\documentclass[10pt,prb,aps,twocolumn,showpacs,floatfix]{revtex4}

\usepackage{graphicx}
\usepackage{amssymb}
\usepackage{color}
\usepackage{amsmath}
\usepackage{amsbsy}
\usepackage{amsmath}
\usepackage{epsfig}
\newcommand\be{\begin{eqnarray}}
\newcommand\ee{\end{eqnarray}}
\newcommand\beq{\begin{equation}}
\newcommand\eeq{\end{equation}}

\newcommand{\non}{\nonumber}

\newcommand{\si}{\sigma}

\newcommand{\vk}{\vec k}

\begin{document}

\title{Entanglement generation in periodically driven integrable systems: dynamical phase transitions and steady state}
\author{Arnab Sen, Sourav Nandy, and K. Sengupta}

\affiliation{Department of Theoretical Physics, Indian Association
for the Cultivation of Science, Jadavpur, Kolkata 700032, India.}

\begin{abstract}

We study a class of periodically driven $d-$dimensional integrable
models and show that after $n$ drive cycles with frequency $\omega$,
pure states with non-area-law entanglement entropy $S_n(l) \sim
l^{\alpha(n,\omega)}$ are generated, where $l$ is the linear
dimension of the subsystem, and $d-1 \le \alpha(n,\omega) \le d$. We
identify and analyze the crossover phenomenon from an area ($S \sim
l^{ d-1}$ for $d\geq1$) to a volume ($S \sim l^{d}$) law and provide
a criterion for their occurrence which constitutes a generalization
of Hastings' theorem to driven integrable systems in one dimension.
We also find that $S_n$ generically decays to $S_{\infty}$ as
$(\omega/n)^{(d+2)/2}$ for fast and $(\omega/n)^{d/2}$ for slow
periodic drives; these two dynamical phases are separated by a
topological transition in the eigensprectrum of the Floquet
Hamiltonian. This dynamical transition manifests itself in the
temporal behavior of all local correlation functions and does not
require a critical point crossing during the drive. We find that
these dynamical phases show a rich re-entrant behavior as a function
of $\omega$ for $d=1$ models, and also discuss the dynamical
transition for $d>1$ models. Finally, we study entanglement
properties of the steady state and show that singular features
(cusps and kinks in $d=1$) appear in $S_{\infty}$ as a function of
$\omega$ whenever there is a crossing of the Floquet bands. We
discuss experiments which can test our theory.

\end{abstract}

\date{\today}

\maketitle

\section{Introduction}
\label{intro}

Entanglement entropy of a correlated many-body system has been the
subject of intense theoretical study in recent years
\cite{rev1,rev2}. It is well-known that this quantity, which probes
non-local correlations of a quantum many-body state, may serve as an
indicator of topological properties of correlated ground states of
several systems such as spin-liquids \cite{rev1}, quantum Hall
systems \cite{qh1}, symmetry broken systems \cite{symbr1},
interacting fermions \cite{fermlit} and topological insulators
\cite{topin1}. In addition, it also contains information about
universal features of quantum phase transitions and serves as an
indicator of a topological quantum phase transition where the
absence of a local order parameter renders the usual Landau
description of the transition impossible \cite{qpt1,qpt2}. More
recently, entanglement entropy for ground states of conformal and/or
large $N$ field theories have received a lot of attention both from
a field theory perspective \cite{conformlit,largenlit} as well as
from the perspective of their gravity duals in an AdS background
\cite{adsrev}.

In most cases, the computation of entanglement entropy involves
computing the reduced density matrix of a quantum many-body system
and/or interacting field theories. This is usually done by starting
from the density matrix corresponding to the quantum ground state,
followed by division of the system into a subsystem of linear
dimension $l$ and "environment" of linear dimension $L-l$, where $L$
is the system size. One then integrates out the degrees of freedom
which belong to the environment and obtains the reduced density
matrix $\rho_r$ for the subsystem. The entanglement entropy can then
be computed from $\rho_r$ by using one of the possible measures
\begin{eqnarray}
S_n &=& (1-n)^{-1} {\rm Tr} [\rho_r^n] \nonumber\\
S &=& -{\rm Tr} [\rho_r \ln \rho_r] = \lim_{n\to 1} S_n
\label{entangdef},
\end{eqnarray}
where $S_n$ is the $n^{\rm th}$ R\'enyi entropy and $S$ denotes the
Von-Neumann entanglement entropy. Such a quantity measures the
entanglement of the subsystem with rest of the system. It is well
known that for a generic short-ranged Hamiltonian, $S$ is controlled
by the boundary between the subsystem and environment leading to
\begin{eqnarray}
S &\sim&  l^{d-1} \quad d \geq 1.
 \label{arealaw1}
\end{eqnarray}
In one dimension, the area law behavior of ground states with local
Hamiltonians and gapped spectrum has been proven and also goes by
the name of Hastings' theorem \cite{hasref1} ($d=1$ critical points
have a further multiplicative logarithmic correction, i.e., $S \sim
\ln l$). It is believed that ground states of local Hamiltonians in
higher dimensions show an analogous area law
behavior~\cite{cirac_etal}. More recently, the subleading terms in
the expression of entanglement entropy have been carefully studied
\cite{rev1}. Such studies yield $S \sim l^{d-1} +\Gamma$ (for
$d>1$). The subleading factor $\Gamma$, when non-zero, encodes the
topological character of the ground state of the system and is often
refereed to as the topological entanglement entropy. We note here
that the boundary-law for the entanglement entropy encoded in Eq.\
\ref{arealaw1} is surprisingly robust. In fact, the only known
violation for an area law for quantum ground states (apart from
critical points in $d=1$) occur for systems with gapless Fermi
surface in $d>1$ where the correlation functions of local operators
become long-ranged leading to $S \sim l^{d-1} \ln l$. However, it is
known that other states in the Hilbert space of a quantum system
which are not the ground state of the local Hamiltonian describing
the system may have non-area law behavior for the entanglement
entropy. A class of these states, obey the volume law, namely, $S
\sim l^d$ for a $d-$ dimensional quantum system.

There have been several studies in recent past on non-equilibrium
dynamics of closed quantum systems \cite{rev3,rev4,rev5,rev6}. Such
studies initially focused on behavior of the system following
quench and ramp dynamics. A class of these studies dealt with the
behavior of such systems following a ramp through a second order
phase transition and discussed the presence/absence of Kibble-Zurek
scaling and their extensions
\cite{rev3,rev4,anatoli1,sengupta1,sen1}. The other class focused
on long-time behavior of these systems following a quench and the
character of the steady states they attain \cite{rigol1}. Such
studies are mainly motivated by presence of experimental platforms
in the form of ultracold atom systems where relevant experiments can
be carried out \cite{rev7}. More recently, the properties of
periodically closed driven quantum systems which involved multiple
passage through an intermediate quantum critical point has received
a lot of attention; in particular such dynamics has been shown to
lead to interesting phenomenon such as dynamic freezing \cite{das1,
sengupta2} and to novel steady states \cite{das2}. Moreover, such
driven systems are known to undergo dynamic phase transitions which
manifests itself in cusp like behavior of the Loschmidt echo and can
be understood as a consequence of the non-analyticities (Fischer
zeroes) of the dynamical free energy of the driven systems
\cite{pol1,sub1,sub2,sangita1}.

The properties of entanglement entropy for states resulting from
non-equilibrium dynamics of closed quantum system has also received
some attention in recent years. The initial studies in this
direction focused on integrable spin-models and on single/two-spin
entanglement \cite{entang1,sen2}. Later, there have been several
studies on the entanglement properties of quantum systems right
after a ramp through a critical point \cite{rampstudy} and on
evolution of entanglement entropy after a sudden quench
\cite{huse1}. The first class of study did not find violation of the
area-law behavior while the second class of study found ballistic
spread of entanglement entropy followed by a plateau where it
attains a constant value at long times. This constant value $S_f$
follows volume law ($S_f \sim l^{d}$) and thus one sees a crossover
from an area to a volume law as a result of the dynamics. Such a
spread has also been studied in the context of integrable spin
models where the system is allowed to evolve after being driven
periodically; an analogous growth of entanglement entropy leading to
a volume law was also observed for such protocols
\cite{peschelquench}. However, to the best of our knowledge, the
fate of the entanglement entropy of a periodically driven quantum
system, where the drive generates multiple passage of the system
through an intermediate quantum critical point, has not been studied
so far. In particular, the crossover of entanglement entropy from an
area to a volume law behavior as a function of drive frequency
$\omega$ and number of drive cycles $n$ has not been investigated in
this context. The convergence of the reduced density matrix of a
subsystem to the final ($n\rightarrow \infty$)
steady state density matrix
has also not been explored for a periodically driven system.

In this work we aim to fill up this gap in the literature by
studying a class of integrable models subjected to a periodic drive
with frequency $\omega$ for $n$ cycles whose Hamiltonian is given by
\begin{eqnarray}
H = \sum_{\vec k} \psi_{\vec k}^{\dagger} \left[ (g(t)-b_{\vec
k})\tau_3 + \Delta_{\vec k} \tau_1 \right] \psi_{\vec k},
\label{hamdef1s}
\end{eqnarray}
where $\vec k$ is $d-$dimensional momentum vector, $\psi_{\vec k} =
(c_{\vec k}, c_{-\vec k}^{\dagger})^T $, $c_{\vec k}$ denotes
fermionic annihilation operator, $\tau_3$ and $\tau_1$ are Pauli
matrices, and $g(t)$ is a periodic function of time. Such
Hamiltonians constitute fermionic representations of Ising and XY
models in $d=1$, the Kitaev model in $d=2$
\cite{subir1,sengupta1,kitaev1,feng1, yao1, nussinov1}, and Dirac
quasiparticles of graphene and atop topological insulator surfaces
\cite{revgraphene,revti}. The relation of Eq.\ \ref{hamdef1s} to
Ising and Kitaev models has been charted out in Appendix
\ref{iskit}. In what follows, we shall study the entanglement
entropy $S_n(l)$ (with $l$ denoting the linear dimension of the
subsystem) of a system described by Eq.\ \ref{hamdef1s} after $n$
drive cycles with frequency $\omega$.

The main results of our study are as follows. First, we find that
for a generic $n$ (and also in the $n \to \infty$ limit where the
system's state is described by the diagonal ensemble) and $\omega$,
$S_n \sim l^{\alpha(n,\omega)}$, where $\alpha(n,\omega)$ satisfies
$d-1 \le \alpha(n, \omega) \le d$; thus a periodic drive may be used
to generate states with non-area-law entanglement entropy in a
controlled manner. We construct a Hamiltonian ${\mathcal H}_t$ for
which the state obtained after $n$ drive cycles is the ground state
and show that the crossover of $S_n$ from an area to a non-area law
can be related to the short-/long-range nature of ${\mathcal H}_t$;
our analysis in this regard constitutes a generalization of
Hastings' theorem to driven $d=1$ integrable quantum systems.
Second, we show that such periodically driven systems show two
distinct dynamical phases when the driving frequency is varied;
$S_n$ relaxes to its steady state value $S_{\infty}$ as
$(\omega/n)^{(d+2)/2} [(\omega/n)^{d/2}]$ in the former [latter]
phase which corresponds to fast [slow] $\omega$. These two phases
are separated by a transition occurring at a critical drive
frequency $\omega_c$ which involves change in topology of spectrum
of the system's Floquet Hamiltonian $H_F$. Third, we show that these
phases exhibit re-entrant behavior as a function of $\omega$ for
$d=1$. We provide a generic phase diagram for this phenomenon as a
function of the drive frequency and amplitude for the 1D Ising
model. We also provide an analytical expression for the phase
boundary for periodic kick protocol which matches the numerical
results accurately. We discuss the nature of this transition in
$d=2$ models and point out some essential differences compared to
the $d=1$ case. We demonstrate that the dynamic phase transition
unraveled here is of fundamentally different origin from the class
of transitions studied in Refs.\
\onlinecite{pol1,sub1,sub2,sangita1} and point out the essential
difference between the two.  Finally, we study $S_{\infty}$, as
obtained from the steady state (diagonal ensemble), as a function of
$\omega$ and show the presence of singularities in $S_{\infty}$
(cusps and kinks in $d=1$) that
are universal features directly related to the crossing of Floquet
bands obtained from the time evolution operator $U(T)$ for one
complete driving period $T$. We discuss experiments which can test
our theory.

The plan of the rest of the work is as follows. In Sec.\
\ref{crosso1}, we numerically demonstrate the area- to volume-law
crossover of $S$ and chart out the construction of $H_t$. This is
followed by Sec.\ \ref{dtrans1}, where we discuss the dynamical
transition reflected in relaxation of $S_n$ and various local
quantities and relate such behavior to the properties of the Floquet
spectrum of the driven system. Next, we discuss the behavior of the
steady state entanglement entropy $S_{\infty}$ as a function of the
drive frequency in Sec.\ \ref{ensteady}. Finally, we discuss our
results, chart out possible experiments which can test them, and
conclude in Sec.\ \ref{diss}. Some details of calculations are shown
in the appendices.

\section{Area- to Volume-Law Crossover}
\label{crosso1}

We begin this section with a brief sketch of our method for
computing $S_n$ numerically. In what follows, we vary $g(t)$ (Eq.\
\ref{hamdef1s}) periodically in time. Although most of our results
would be protocol independent, for numerical purposes, unless
mentioned otherwise, we use the square pulse drive protocol:
\begin{eqnarray}
g(t)&=& g_i, \quad (n-1)T \le t\le (n-1/2)T \nonumber\\
&=& g_f, \quad (n-1/2)T \le t\le n T,\label{sqpro}
\end{eqnarray}
where $T=2\pi/\omega$ is the time period. The issue of protocol
independence of some of our results is charted out in further detail
in Appendix \ref{proin}. To solve the dynamics, we define the annihilation
operators $\gamma_k(t)$:
\begin{eqnarray}
\gamma_{\vec k} &=& u_{\vec k}(t) c_{\vec k} + v^{\ast}_{\vec k}(t)
c_{-\vec k}^{\dagger}. \label{gamdef}
\end{eqnarray}
Here $u_{\vec k}(t)$ and $v_{\vec k}(t)$ satisfy the Sch\"rodinger
equation
\begin{eqnarray}
i \partial_t |\psi_{\vec k}\rangle = H_{\vec k}(t) |\psi_{\vec
k}\rangle,  \label{sch1}
\end{eqnarray}
where $|\psi_{\vec k}\rangle= (u_{\vec k}, v_{\vec k})^T$ and we
have set $\hbar=1$. The wavefunction $|\psi (t) \rangle$ of the 
entire system equals
\be
|\psi (t) \rangle &=& \otimes_{\vec k \in {\rm BZ}/2}|\psi_{\vec{k}}(t) \rangle \nonumber \\
|\psi_{\vec{k}}(t) \rangle &=& u_{\vec{k}}(t)c^{\dagger}_{\vec{k}} c^{\dagger}_{-\vec{k}} |0 \rangle + v_{\vec{k}}(t)|0 \rangle 
\label{fullwavef} 
\ee
where $\vec{k}$ is taken over half of the Brillouin zone (BZ) and 
$|0 \rangle$ denotes the vacuum of the $c$ fermions. 

Having obtained $|\psi_{\vec k}(nT) \rangle$, the calculation of
$S_n$ requires the construction of two $l^d \times l^d$ matrices
\cite{pes1}, ${\bf C}$ and ${\bf F}$, whose elements can be
constructed by knowing $u_k(t)$ and $v_k(t)$ after $n$ drive period:
\begin{eqnarray} C_{ij} &=& \langle c^{\dagger}_{\vec i} c_{\vec j}
\rangle_n =
2 \sum_{\vec k \in {\rm BZ}/2} |u_{\vec k}(t)|^2 \cos(\vec k \cdot (\vec i- \vec j))/L^d \label{matrices1}  \nonumber \\
F_{ij} &=& \langle c^{\dagger}_{\vec i} c^{\dagger}_{\vec j}
\rangle_n = 2 \sum_{\vec k \in {\rm BZ}/2} u^*_{\vec k}(t) v_{\vec
k}(t) \sin(\vec k \cdot (\vec i-\vec j))/L^d \nonumber \\
 \end{eqnarray}
%%where the $\vec{k}$ summation is taken over half the Brillouin zone
%%(BZ) and 
where $i,j$ refer to sites in the subsystem.
%The details of these
%calculation can be found in App.\ \ref{secmeasure}.
Using these
expressions, we construct the $2l \times 2l$ matrix ${\mathcal
C}_n(l)$ given by
\begin{eqnarray}
{\mathcal C}_n(l) &=& \left( \begin{array}{cc} \mathbf{I-C} &
\mathbf{F} \\ \mathbf{F}^{\ast} & \mathbf{C } \end{array} \right).
\label{smatrix}
\end{eqnarray}
$S_n$ can then be obtained from $2l$ eigenvalues $p_i$ of ${\mathcal
C}_n(l)$: $S_n(l) = -{\rm Tr}[\rho_r \ln \rho_r]= -\sum_{i=1}^{2l} p_i
\log(p_i)$, where $\rho_r$ is the subsystem density matrix after $n$
drive cycles \cite{pes1}. The details of this calculation is
sketched in Appendix \ref{pesca}.

\begin{figure}
{\includegraphics[width=\hsize]{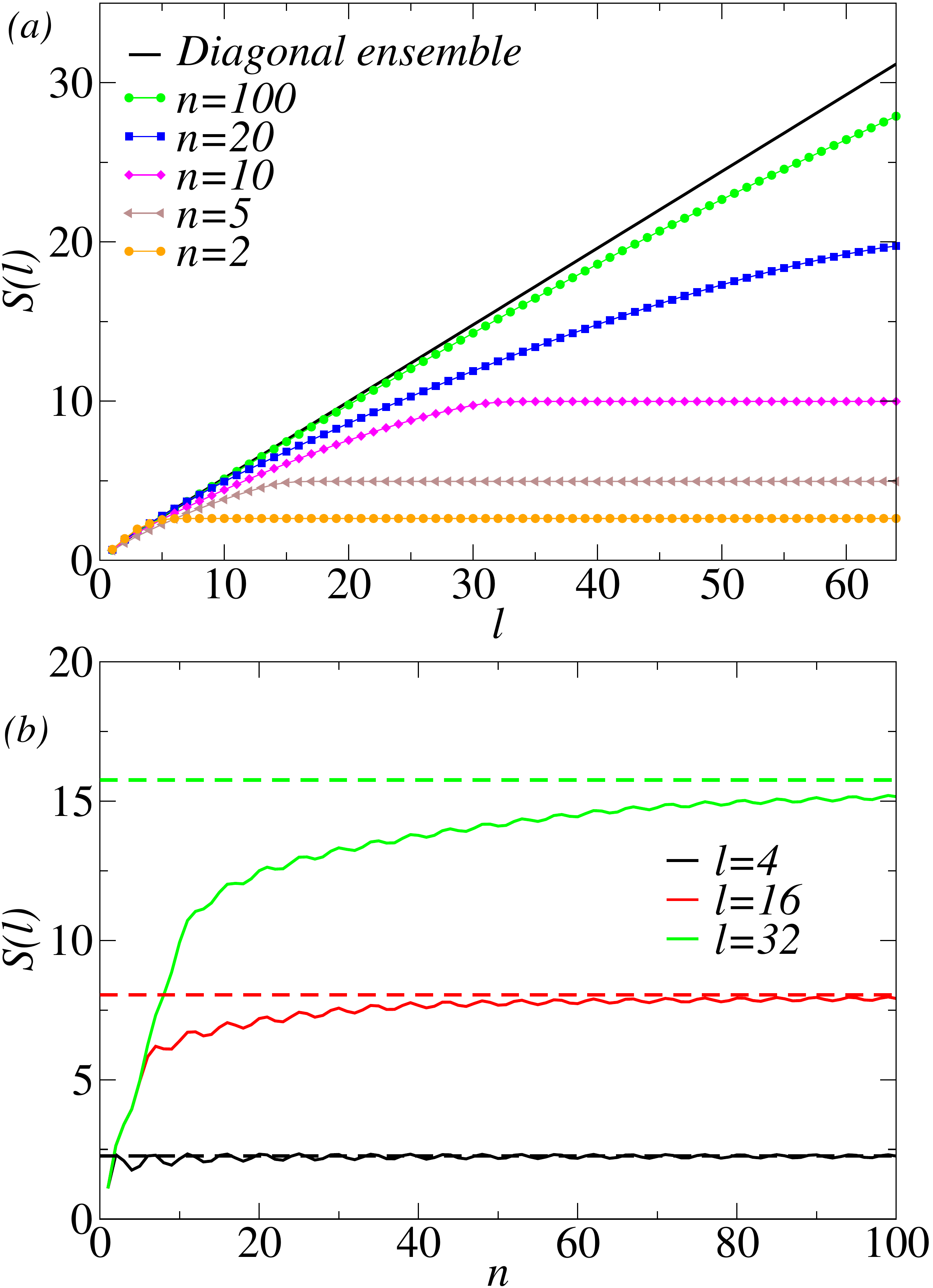}} \caption{ $S_n$ for 1D Ising
model for which  $\Delta_k=\sin(k)$, $b_k = \cos(k)$ and $J=1$. The
transverse field $h(t)=g(t)$ follows the square pulse protocol
($g_i=2$ and $g_f=0$), and the ground state at $g_i=2$
is taken to be the starting state.
(a) $S_n$ versus $l$ for several $n$ and
$\omega=\pi$ (b) $S_n$ versus $n$ for several $l$ and $\omega=\pi$.
\label{fig1}}
\end{figure}

The result of such a numerical study is shown in Fig.\ \ref{fig1}. In
Fig.\ \ref{fig1}(a), we plot $S_n(l)$ as a function of $l$ for $d=1$
Ising model and several $n$. From this plot, we find that the
minimum value of $l$ beyond which $S_n(l)$ satisfies the area law
(i.e., $S_n(l) \sim$ constant in $d=1$) diverges as $n \to \infty$ leading to
genuine non-area scaling. In this regime, $S_n(l) \sim
l^{\alpha(n,\omega)}$ where $d-1 \le \alpha(n, \omega) \le d$.
The $n \rightarrow \infty$ result is reproduced by the Diagonal Ensemble;
and we detail
those calculations in Sec.~\ref{ensteady}. Next,
as shown in Fig.\ \ref{fig1}(b), we find that $S_n(l)$ grows
linearly as a function of $n$ for a fixed $l$ and then attains
a constant ($l$ dependent) value;
this behavior is qualitatively similar to the linear spread of $S$
following a quench \cite{huse1}.

\begin{figure}
{\includegraphics[width=\hsize]{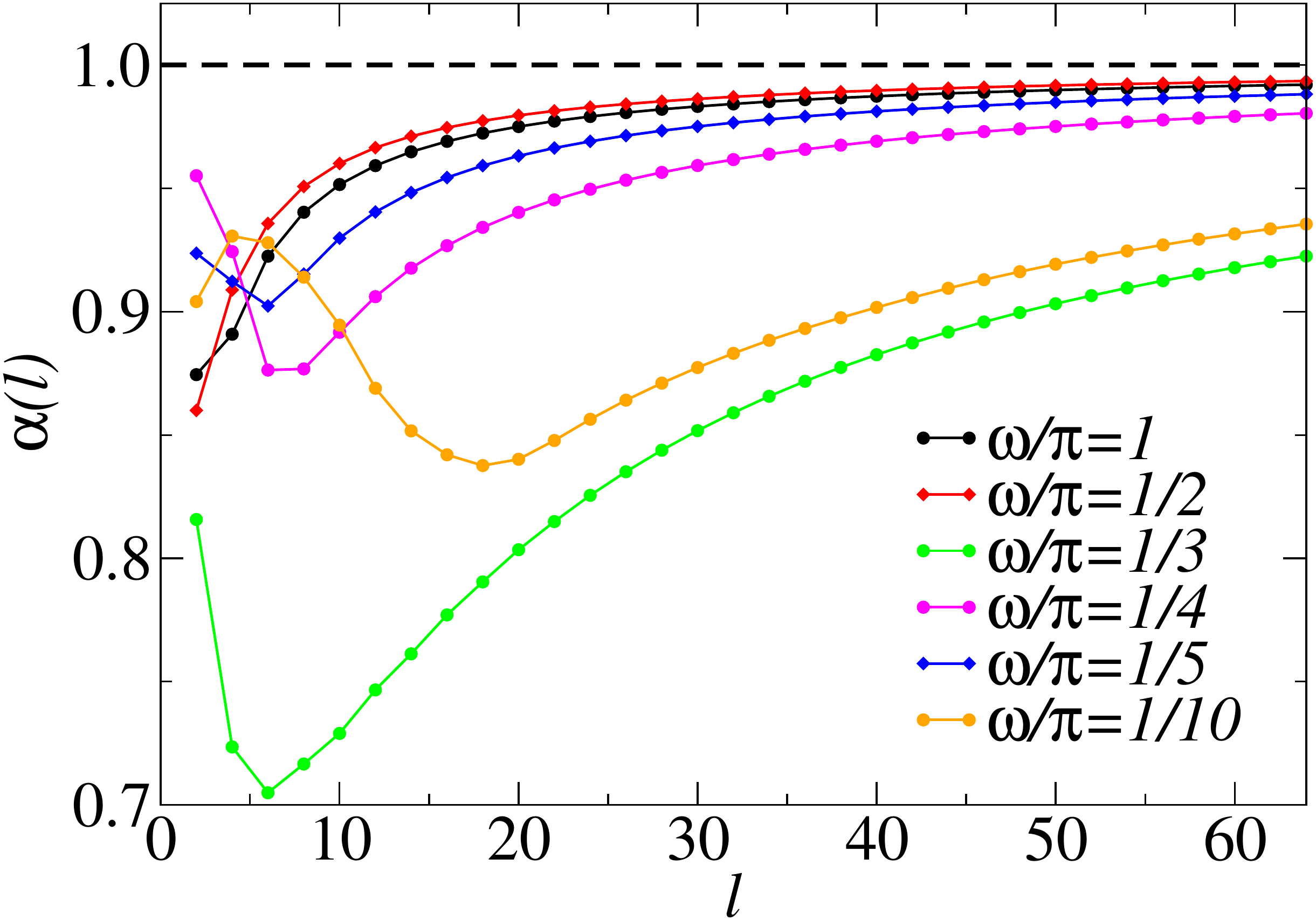}} \caption{ Behavior of
$\alpha(l)$ versus $l$ (Eq.~\ref{esti1}) in the steady state
for several $\omega$ (square pulse protocol with $g_i=2$ and
$g_f=0$) for the 1D Ising model.
\label{fig2}}
\end{figure}

To understand how fast the volume law is approached in the
steady state (Diagonal ensemble) for a certain
drive protocol for different $l$, we define the estimator
\begin{eqnarray}
\alpha(l) = \log[S_{\infty}(2l)/S_{\infty}(l)]/\log(2),
\label{esti1}
\end{eqnarray}
and plot it as a function of $l$ for several representative $\omega$
in Fig.\ \ref{fig2}. In 1D, $\alpha \rightarrow 0(1)$ if
area(volume)-law is satisfied. We find a rapid and monotonic
convergence of $\alpha$ to unity for large $\omega$. In contrast,
for small $\omega$, this convergence is quite slow and has
non-monotonic features. Thus, periodic drives at small $\omega$
offer a route to stabilizing pure quantum states with non-area and
non-volume scaling of $S_n$ \cite{comment2d}.

These results lead to the natural question regarding the area-law
behavior for a finite $l$ and $n$; in particular, it seems desirable
to have an analytical criterion which may, at least qualitatively
predict that area or non-area law behavior of $S$ as a function of
$l$ for a given $n$. We attempt to provide such a criterion below.
To this end, we construct $\mathcal{H}_{{\vec k} t}$ for which $|\psi_{\vec
k}(nT)\rangle$ generated after $n$ drive cycles is the ground state.
The motivation for doing this is as follows. From Hastings' theorem,
the ground state of a local Hamiltonian in one
dimension yields $S$ which exhibits
area law \cite{hasref1}. For a driven integrable system,
$S_{\infty}$ is expected to be described by a
Generalized Gibbs Ensemble (GGE), and thus to
follow volume-law \cite{huse1}. Thus after $n$ cycles of the drive,
for (small)large $n$, one expects $\mathcal{H}_{\vec k t}$ to be
short(long) ranged and the crossover between short- to long-range
behavior of ${\mathcal H}_{{\vec k} t}$ for a given $l$ might
provide an indication of the numerically observed area- to
volume-law crossover of $S_n$.

\begin{figure}
{\includegraphics[width=\hsize]{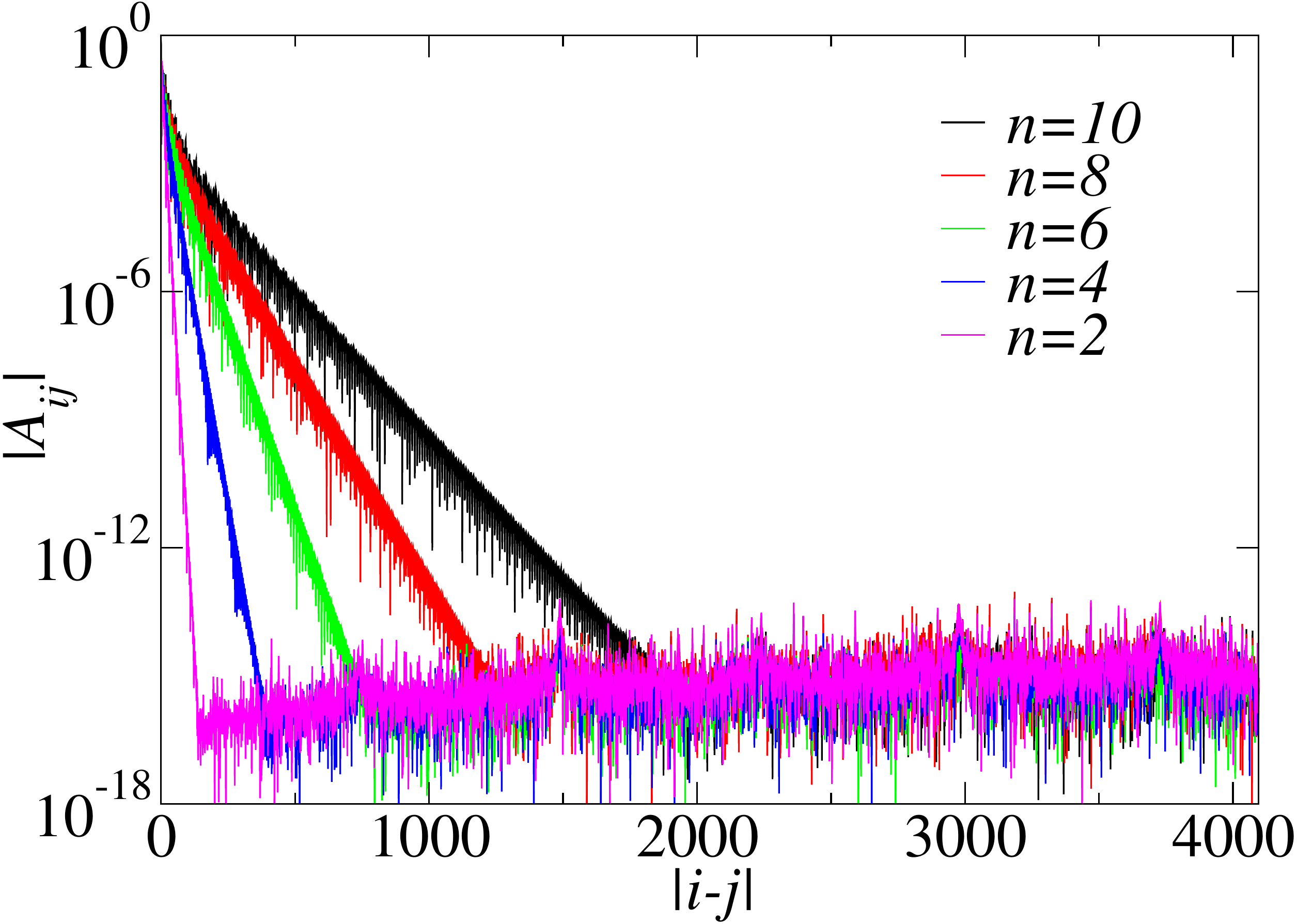}} \caption{$|A_{ij}|$ versus
$r= |i-j|$ for several $n$ and $\omega=2 \pi$ for 1D ising model
where the transverse field $h(t)=g(t)$ follows the square pulse protocol
($g_i=2$ and $g_f=0$).
\label{fig3}}
\end{figure}

To construct $\mathcal{H}_{\vec k t}$ we start from $\psi_k(t_f=nT)$
and seek a solution of
\begin{eqnarray}
{\mathcal H}_{{\vec k} t} = \epsilon_{{\vec k} t} \tau_3 +
\Delta_{{\vec k} t} \tau^+ + \Delta^*_{{\vec k} t} \tau^-,
\label{ht1}
\end{eqnarray}
which satisfies
\begin{eqnarray}
{\mathcal H}_{\vec k t} \psi_{\vec k}(t_f) = -(\epsilon_{{\vec k}
t}^2 + | \Delta_{{\vec k} t}|^2 )^{1/2} \psi_{\vec k}(t_f).
\end{eqnarray}
Assuming that ${\mathcal H}_{\vec k t}  \simeq H_{\vec k}$ in the
adiabatic limit, we find, after some straightforward algebra,
\begin{eqnarray}
\epsilon_{{\vec k} t} &=& \Delta_{\vec k} (|u_{\vec k}(t_f)|^2 -
|v_{\vec k}(t_f)|^2)/(2|u_{\vec k}(t_f)||v_{\vec k}(t_f)|)
\nonumber\\
\Delta_{{\vec k} t} &=& \Delta_{\vec k} \exp(i(\alpha_{\vec k} -
\beta_{\vec k}))
 \label{htcons1}
\end{eqnarray}
where we have defined $\alpha_{\vec k}(\beta_{\vec k}) = {\rm Arg} [
u_{\vec k}(t_f)(v_{\vec k}(t_f))]$. The real-space form of the
effective Hamiltonian is given by Fourier transform of ${\mathcal
H}_{\vec k t}$
\begin{eqnarray}
{\mathcal H}_t = \sum_{\vec i \vec j} (A_{\vec i -\vec j} c_{\vec
i}^{\dagger} c_{\vec j} + B_{\vec i -\vec j} c_{\vec i} c_{\vec j} +
{\rm h.c.}),
\end{eqnarray}
where $A_{\vec i \vec j}$ and $B_{\vec i \vec j}$ are Fourier
transforms of $\epsilon_{\vec k t}$ and $\Delta_{\vec k t}$
respectively. A plot of $|A_{\ij}|$ as a function of $|i-j|$ for the
$d=1$ Ising model (Fig.\ \ref{fig3}), shows $ A_{ij} \sim
\exp[-|i-j|/R_t(n,\omega)]$; this indicates that ${\cal H}_t$
appears short-ranged in the length scale $l\gg R_t(n,\omega)$ and long-ranged
for $l\ll R_t(n,\omega)$. As shown in Fig.\ \ref{fig3}, 
$R_t(n,\omega)$ increases
rapidly with $n$; we find numerically that for $l \gg (\ll)
R_t(n,\omega)$, $S_n$ follows area(non-area)-law in accordance with
Hastings' theorem. This result constitutes a generalization of
Hastings' theorem for driven integrable models.

Before ending this section, we note that the deduction of nature of
$S$ from that of ${\mathcal H}_t$ involves two scales. The first
constitutes the minimal subsystem size $l_{\rm min}$ below which the
subsystem shows a non-area law behavior for a given $n$ while the
second is $R_t(n,\omega)$ which determines the effective long-raged
or short-ranged nature of ${\mathcal H}_t$ with respect to $l$. We
find numerically that $R_t(n,\omega)$ shows a much faster growth with $n$ than
$l_{\rm min}$; thus whereas we do find that $R_t(n,\omega) \gg l$ and 
$R_t(n,\omega) \ll
l$ correspond to two different behaviors of $S(l)$ as a function of
$l$, it is generally not possible to identify $R_t(n,\omega) \sim l$ as the
point of crossover of $S(l)$ from one type of behavior to the other
type.

\section{Dynamical Transition and Phase Diagram}
\label{dtrans1}

In this section we first study the relaxation behavior of the local
quantities in the system to their steady state values and unravel a
dynamic transition in this relaxation behavior as a function of
drive frequency $\omega$ in Sec.\ \ref{dyntra} for both $d=1$ (Sec.\
\ref{1dstuff}) and $d=2$ (Sec.\ \ref{2dstuff}). This is followed by
numerical and analytical study of the corresponding phase diagram
for 1D transverse field Ising model for square and kicked pulse
protocol in Sec.\ \ref{dyntrb} to illustrate the rich re-entrant
behavior as a function of $\omega$ in $d=1$.

\subsection{Relaxation of $S_n$}
\label{dyntra}

\begin{figure}
{\includegraphics[width=\hsize]{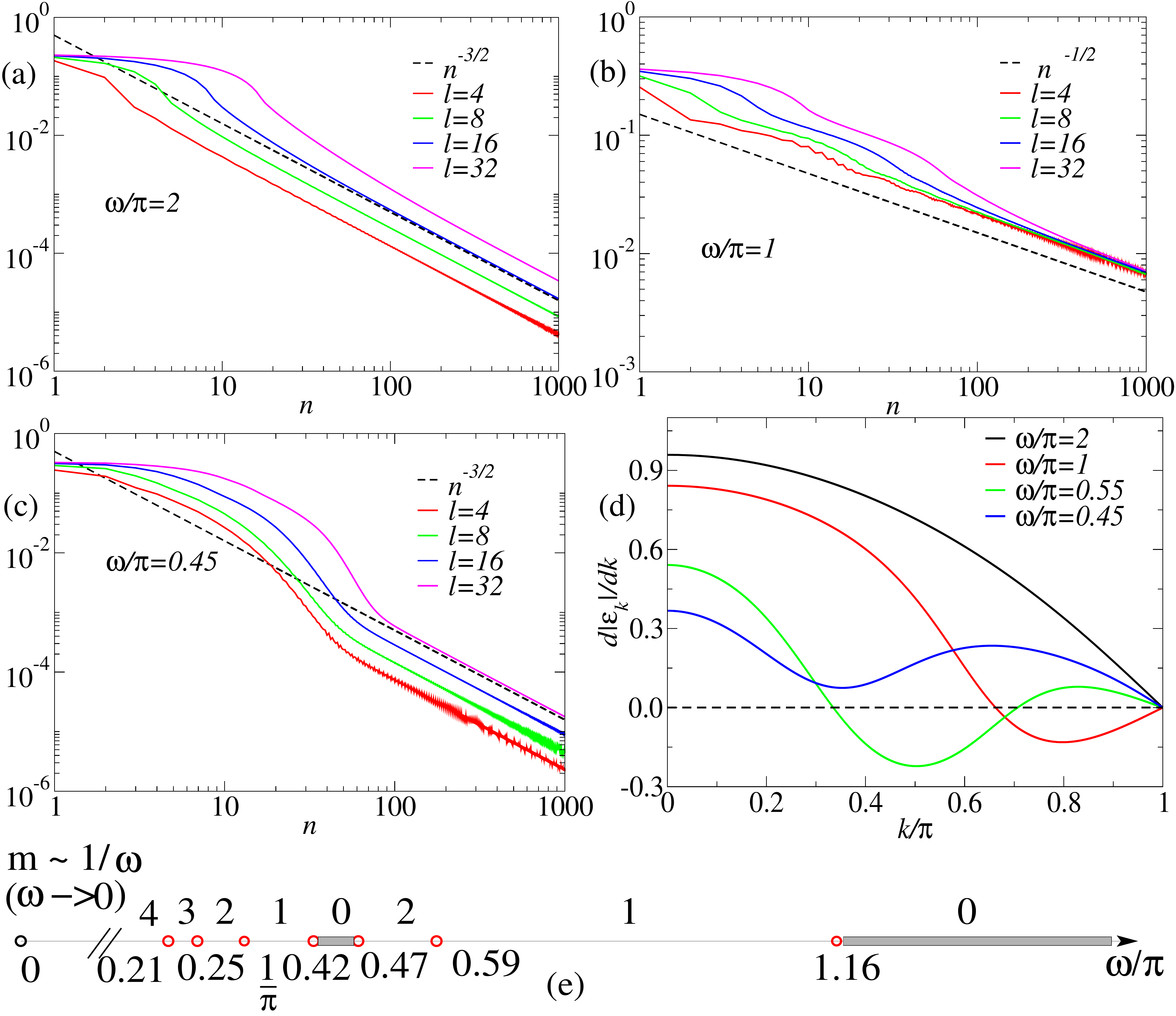}} \caption{${\mathcal D}$
versus $n$ for several $l$ for 1D Ising model and for (a) $\omega= 2
\pi$, (b) $\omega= \pi$, and (c) $\omega=0.45 \pi$. (d) Plot of
$d|\epsilon_k|/dk$ versus $k$ for several $\omega$. (e) Sketch of
the of number of zeroes ($m$) of $d |\epsilon_k|/dk$ with $k \in (0,
\pi)$ as a function of $\omega$ showing re-entrant behavior. The
thick (thin) regions indicate regimes with $n^{-3/2} (n^{-1/2})$
decay. All parameters are same as in Fig.\ \ref{fig1}.
\label{fig4}}
\end{figure}

To study the relaxation of $S_n$ to $S_{\infty}$, we define a
distance measure which provides us information regarding this
relaxation as a function of $n$, and also of the relaxation of the
local density matrix $\rho_r$ (and hence all local quantities
within that subsystem) to the final Diagonal Ensemble result.
For integrable models $S_n$ is
determined by the two-point correlators $\mathcal{C}_n(l)$; thus it
is natural to define \cite{essler1}
\begin{eqnarray}
{\mathcal D} = {\rm Tr}
[(\mathcal{C}_\infty(l)-\mathcal{C}_n(l))^\dagger
(\mathcal{C}_\infty(l)-\mathcal{C}_n(l))]^{1/2}/(2l).
\end{eqnarray}
We note that $0 \le {\mathcal D} \le 1$ and it vanishes only if $C_n
= C_{\infty}$. The details of its calculation is given in Appendix
\ref{secmeasure}. We find numerically for the square pulse protocol
(Figs.\ \ref{fig4} (a), (b), and (c) and \ref{fig5}(a)) that for
both $d=1$ Ising and $d=2$ Kitaev models, ${\mathcal D}$ exhibits
two distinct behaviors corresponding to different dynamical regimes:
${\mathcal D} \sim (\omega/n)^{(d+2)/2}[(\omega/n)^{d/2}]$ in these
two regimes. For the drive amplitude used in Fig.\ \ref{fig4}
we also find that these two dynamical regimes are
separated by re-entrant transitions at $\omega_c =1.16 \pi, \, 0.47
\pi$, and $0.42 \pi$ for the $d=1$ Ising model;
however, for $d=2$
Kitaev model, there is a single such transition $\omega_c= 4.01\pi$
\cite{comment4}.

To understand the origin of this transition, we analyze the Floquet
Hamiltonian $H_F$ for the driven system. After $n$ drive cycles, the
wavefunction is given by
\begin{eqnarray}
|\psi_{\vec k}(t=nT)\rangle &=& U_{\vec k}^n |\psi_{\vec
k}(t=0)\rangle \nonumber\\
&=& \exp[-in H_{\vec k F} T] |\psi(t=0)\rangle,
\end{eqnarray}
where $H_{\vec k F}$ is the Floquet Hamiltonian of the system for
the wavevector $\vec k$ \cite{luca1}, $H_F= \sum_{\vec k} H_{\vec k
F}$, and $U_{\vec k}$ is given by
\begin{eqnarray}
U_{\vec k} = \cos(\theta_{\vec k}) \exp[i \alpha_{\vec k} \tau_3] -i
\tau_2 \sin(\theta_{\vec k}) \exp[i \gamma_{\vec k} \tau_3].
\label{uexp}
\end{eqnarray}
The parameters $\theta_{\vec k}$, $\alpha_{\vec k}$ and
$\gamma_{\vec k}$ can be expressed in terms of the initial and final
wavefunctions; for example for $|\psi_{\vec k}(t=0)\rangle =
(0,1)^{T}$ and $|\psi_{\vec k}(t=T)\rangle = (u_{\vec k f},v_{\vec k
f})^{T}$, one has $\sin(\theta_{\vec k})= |u_{\vec k f}|$,
$\alpha_{\vec k} = -{\rm Arg} (v_{\vec k f})$ and $\gamma_{\vec k} =
{\rm Arg} (u_{\vec k f})$. We further note that $U_{\vec k}$ becomes
diagonal at the edge and center of the BZ where $\Delta_{\vec k}=0$
leading to $\sin(\theta_{\vec k})=0$ at these points. More details
of these calculations can be found in Appendix \ref{flcons}.

To obtain $H_{\vec k F}$, we note that the unitary nature of
$U_{\vec k}$ guarantees that $H_{\vec k F}$ can be expressed in
terms of the Pauli matrices. This allows us to write
\begin{eqnarray}
H_{\vec k F} &=& \vec \sigma \cdot \vec \epsilon_{\vec k} = |\vec
\epsilon_{\vec k}| \vec \sigma \cdot \hat n_{\vec k}
\end{eqnarray}
where $\vec \epsilon_{\vec k} = (\epsilon_{\vec k 1}, \epsilon_{\vec
k 2}, \epsilon_{\vec k 3})$, and ${\hat n}_{\vec k i} =
\epsilon_{\vec k i}/|\vec \epsilon_{\vec k}|$. Here the
quasienergies $\epsilon_{\vec k i}$ are given by
\begin{eqnarray}
\epsilon_{1\vec k} &=& -|\vec \epsilon_{\vec k}| \sin(\theta_{\vec
k}) \sin(\gamma_{\vec k})\sin(T|\epsilon_{\vec
k}|)/D_{\vec k}
\nonumber\\
\epsilon_{2 \vec k} &=& -|\vec \epsilon_{\vec k}|\sin(\theta_{\vec
k}) \cos(\gamma_{\vec k})\sin(T|\epsilon_{\vec
k}|)/D_{\vec k}
\nonumber\\
\epsilon_{3 \vec k} &=& -|\vec \epsilon_{\vec k}| \cos(\theta_{\vec
k}) \sin(\alpha_{\vec k})\sin(T|\epsilon_{\vec
k}|)/D_{\vec k}
\nonumber\\
D_{\vec k}&=& \sqrt{1-\cos^2(\theta_{\vec k}) \cos^2(\alpha_{\vec
k})}.
\label{ueq3a}
\end{eqnarray}
Thus one can write
\begin{eqnarray}
U_{\vec k} &=& \exp[-i(\vec \sigma \cdot \hat n_{\vec k}) \phi_{\vec
k}],\label{ueq3}
\end{eqnarray}
where $\phi_{\vec k} = T|\vec \epsilon_{\vec k}|$. We work in the
reduced zone scheme where $\phi_{\vec{k}} \in [0,\pi]$ and each of
the component of  $\vec{\epsilon}_{\vec{k}}$ is restricted to
$[-\pi/T,\pi/T]$. We note that from the above structure it is clear
that $n_{1 \vec k}=n_{2 \vec k}=0$ and $n_{3 \vec k}=\pm 1$ at the
edge and center of the BZ where $U_{\vec k}$ is diagonal; at all
other point in the BZ, all three components of $\vec n_{\vec k}$ are
generally non-zero. Using Eq.\ \ref{uexp}, one can then obtain the
Floquet spectrum
\begin{eqnarray}
|\vec \epsilon_{\vec k}| &=&   \arccos[\cos(\theta_{\vec k})
\cos(\alpha_{\vec k})]/T.   \label{ueq3as}
\end{eqnarray}
The details of calculation of the Floquet spectrum is given in Appendix
\ref{flcons}.

Having obtained an expression for the Floquet spectrum, we now
express the elements of $\mathcal{C}_n(l)$ (Eqs.\ \ref{smatrix} and
\ref{matrices1}) in terms of parameters of $U_{\vec k}$. A
straightforward calculation (see Appendix \ref{secmeasure} for details)
yields, for $L \rightarrow \infty$ and $|\psi(t=0)\rangle_{\vec
k}=(0,1)^T$, where $\delta C_{ij}(n)= \langle c_{\vec i}^{\dagger}
c_{\vec j} \rangle_n - \langle c_{\vec i}^{\dagger} c_{\vec j}
\rangle_{\infty}$ and similarly for $\delta F_{ij}(n)$:
\begin{eqnarray}
\delta C_{ij}(n) &=&    \int \frac{d^d k}{(2\pi)^d} f_1(\vec{k})\cos(2n\phi_{\vec k}) \label{generalresults} \\
 \delta F_{ij}(n) &=&  \int \frac{d^d k}{(2\pi)^d} (f_2(\vec{k})\cos(2n\phi_{\vec k}) +f_3(\vec{k})\sin(2n\phi_{\vec k})) \nonumber
\end{eqnarray}
with
\begin{eqnarray}
f_1(\vec{k}) &=& -(1-\hat{n}^2_{\vec{k}3})\cos(\vec{k}\cdot
(\vec{i}-\vec{j})), \,\, f_2(\vec{k}) = -i \hat{n}_{\vec{k}3}
f_3(\vec{k}) \nonumber\\
f_3(\vec{k}) &=& i({n}_{\vec{k}1}+i{n}_{\vec{k}2})\sin(\vec{k}\cdot
(\vec{i}-\vec{j})). \label{ffns}
\end{eqnarray}
Importantly, $f_{1,2,3} (\vec{k})$ all
vanish at the BZ edges and center.

It is clear from Eq.\ \ref{generalresults} that for large $n$, the
dominant contributions to the relaxation behavior comes from the
stationary points of $\phi_{\vec k}$: $d|\vec \epsilon_{\vec
k}|/dk_i=0$. Using Eq.\ \ref{ueq3as}, we find that such stationary
points occur if either
\begin{eqnarray}
\cot(\theta_{\vec k}) d \alpha_{\vec k}/dk_i = - \cot(\alpha_{\vec
k}) d \theta_{\vec k}/dk_i \label{cond1b}
\end{eqnarray}
or $\sin(\theta_{\vec k})=0= d \alpha_{\vec k}/dk_i$. We note that
Eq.\ \ref{cond1b} holds for any protocol; the protocol details
appear in the expression of $\theta_{\vec k }$ and $\alpha_{\vec k
}$ without altering its form.

\subsection{Approach to steady state in $d=1$}
\label{1dstuff}

For $d=1$ models, when $\omega \gg 1$, one can approximate $H_{k F}
\sim \bar{H}(k)$, where $\bar{H}$ denotes the time-averaged
Hamiltonian over one cycle, by using $1/\omega$ as a perturbation
parameter in the Dyson series for $U_k$. In this limit, there are
two stationary points at $k=0,\pi$ and thus $\sin(\theta_{\vec
k})=0= d \alpha_{\vec k}/dk_i$ is satisfied. As $\omega$ is
decreased below $\omega_c$, an additional stationary point emerges
at $k = k_0 \in (0,\pi)$ which satisfies Eq.\ \ref{cond1b}(Fig.\
\ref{fig4}(d)). This leads to a qualitative change in the relaxation
properties of the matrix elements (Eq.\ \ref{generalresults}) which
can be understood as follows. The contribution of a saddle point at
$k=k_0$ to Eq. \ref{generalresults} can be estimated to be
\begin{eqnarray}
&&\int f_i(\vec k) \exp(in\phi(\vec k))d^dk \approx
\exp(in\phi(\vec k_0))(n |\phi^{''}(\vec k_0|))^{-d/2}\nonumber\\
&& \times \exp(\pi i \mu/4) \left(f_i(\vec k_0)+i\frac{f_i^{''}(\vec
k_0)}{2\phi^{''}(\vec k_0)}\frac{1}{n}+\mathcal{O}(1/n^2) \right)
\label{sadpont1}
\end{eqnarray}
where $\mu$ is the sign of $\phi^{''}(\vec k_0)$, and $f_i(\vec k)$
are smooth functions around $\vec k=\vec k_0$. For $d=1$, at $k=0,
\pi$, $\sin(\theta_k)=0$, ${\it i.e.}$,
$\hat{n}_{k1},\hat{n}_{k2}=0$ and $\hat{n}_{k3}=\pm 1$ leading to
$f_i(k_0)=0$. If these happen to be the only zeroes of $d
|\epsilon_{k}|/dk$, all elements of ${\mathcal C}_n(l)$ (and hence
${\mathcal D}$) receive first non-zero contribution from the
$f''(k)$ term in Eq.\ \ref{sadpont1} leading to a convergence  to
the GGE as $(\omega/n)^{3/2}$. However, for a smaller $\omega$ the
contribution from stationary point at $k =k_0 \ne 0, \pi$ (where
$f_i(k_0)\ne 0$) changes the relaxation behavior of ${\mathcal D}$
to $(\omega/n)^{1/2}$(Eq.\ \ref{sadpont1}). The appearance of such a
new zero constitute a change in topology of spectrum of $H_F$. In
addition, $\omega_c$ is expected to be finite in general since the
number of zeroes of $d|\epsilon_k|/dk$ cannot change continuously
with $\omega$. In fact, this number cannot change {\it perturbatively}
in $1/\omega$ and hence, the dynamical transitions we discuss here
are beyond a Magnus expansion treatment of the Floquet Hamiltonian (for a
review of which, see Ref.~\onlinecite{pol1}) to
any order.

Indeed for the square pulse protocol it can be shown
the first zero of $d
|\epsilon_{k}|/dk$ where $k \neq 0,\pi$ 
appears at a finite $\omega_c$ value (details given in Appendix.~\ref{flcons}) which satisfies
\begin{widetext}
\begin{eqnarray}
(g_i+g_f+2g_ig_f)T_0 \sin\left( \left(2+g_i+g_f \right)T_0 \right)+
\frac{(g_f-g_i)^2 \sin \left((1+g_i)T_0 \right) \sin \left((1+g_f)T_0 \right)}{(1+g_i)(1+g_f)} = 0
\label{condtrans}
\end{eqnarray}
\end{widetext}
where $T_0 = \pi/\omega_c$ is the first non-zero solution of
Eq.~\ref{condtrans} given $g_i$ and $g_f$. Hence our result
constitutes an example of change in relaxation behavior of {\it any
correlation function} of a periodically driven integrable
Hamiltonian due to change of topology of their $H_F$
\cite{comment4}. We note that the transition unraveled in this work
is of fundamentally different origin from dynamical transitions
discussed in Refs.\ \onlinecite{pol1,sub1, sub2, sangita1}; in
contrast to these transitions, the present one leaves its imprint on
the temporal behavior of {\it all} local correlation functions.

As $\omega$ is decreased, the number of zeroes of $d
|\epsilon_k|/dk$, $m$, between $0<k<\pi$ changes. Such a change is
non-monotonic in nature for large $\omega$ (Fig.\ \ref{fig4}(c) and
(d)) where $m$ is small. It is thus possible that in some frequency
range $m$ may revert back to zero leading to re-entrant behavior;
numerically, for square pulse protocol, we find that this occurs at
$\omega_c= 0.47 \pi$ for $g_i=2,g_f=0$ (Fig.\ \ref{fig4}(e)). As
$\omega$ is further decreased, $m$ becomes finite at $\omega = 0.42
\pi$ and continue to increase monotonically with decreasing
$\omega$: in fact $m \sim \omega^{-1}$ for small $\omega$ (Fig.\
\ref{fig4}(e)) (and this is a generic feature~\cite{comment4}), thus
ruling out re-entrance here and leading to a $(\omega/n)^{1/2}$
convergence to steady state as $\omega \rightarrow 0$. The phase
diagram for the two dynamical regimes has a rich structure as a
function of the amplitude and frequency of the periodic drive due to
the re-entrance effects and this will be worked out in more detail
in Sec.~\ref{dyntrb}.

\subsection{Approach to steady state in $d>1$}
\label{2dstuff}
For $d>1$, we note that for large $\omega$, the condition
$\sin(\theta_{\vec k})=0 = d\alpha_{\vec k}/dk_i$ is expected to be
satisfied leading to ${\mathcal D} \sim (\omega/n)^{(d+2)/2}$. As
$\omega$ is decreased, new zeros of $d|\vec \epsilon_{\vec k}|/dk_i$
are expected to appear at $\vec k= \vec k_0$ which satisfies Eq.\
\ref{cond1b}. Generically, one expects such solutions to constitute
discrete point(s) in the Brillouin zone or there may be no solutions
at all. In the former case, one would find a transition to
$(\omega/n)^{d/2}$ scaling (Eq.\ \ref{sadpont1}) along with possible
re-entrant behavior similar to the $d=1$ model.

\begin{figure}
{\includegraphics[width=\hsize]{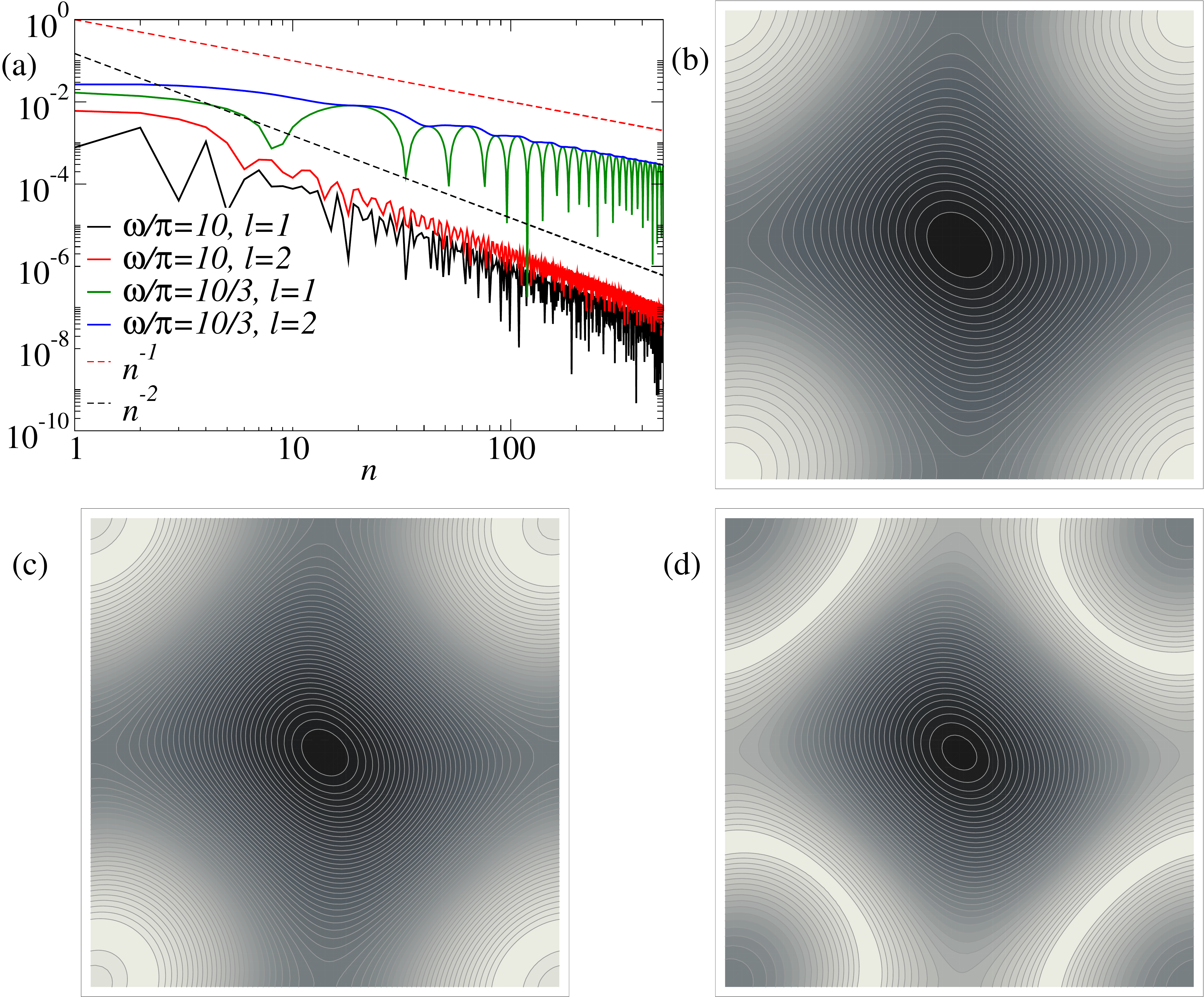}} \caption{(a) ${\mathcal D}$
versus $n$ for 2D Kitaev model for $l=1,2$ with $J_1=J_2=1$,
$g_i=J_3(0)=5$ and $g_f=J_3(T/2)=4$. (b), (c), and (d): Contours of
$|\vec \epsilon_{\vec k}|$ versus $(k_x,k_y)$ for $\omega= 10\pi$
(b), $4 \pi$ (c) and $3.3 \pi$ (d) indicating a transition at
$\omega_c \simeq 4 \pi$; white (black) denotes high (low) values.
 \label{fig5}}
\end{figure}

However, for a class of 2D models, including the Kitaev model, the
existence of a special symmetry leads to solution of Eq.\
\ref{cond1b} along a line(s) in the Brillouin zone. For the Kitaev
model, this can be understood from the fact that the Hamiltonain
(for which $b_{\vec k}=-[J_1\cos(k_x) + J_2\cos(k_y)]$,
$\Delta_{\vec k} = J_1\sin(k_x)+J_2\sin(k_y)$ and $g(t)=J_3(t)$
remain invariant under the simultaneous transformations $J_1
\leftrightarrow J_2$ and $k_x \leftrightarrow k_y$ . The dynamics
preserves this symmetry; consequently the zeroes of $\partial |\vec
\epsilon_{\vec k}|/\partial k_x$ and $\partial |\vec \epsilon_{\vec
k}|/\partial k_y$ coincide and form a 1D curve in the 2D Brillouin
zone leading to a line of zeroes. This phenomenon is illustrated in
Fig.\ \ref{fig5}(b) (c) and (d). The additional line of zeroes
appears when $\omega$ is changed from $10\pi$ (Fig. \ref{fig5}(b))
to $3.3 \pi  $ (Fig.\ \ref{fig5}(d)). The critical point, where the
line of zeroes first appear is $\omega_c \simeq 4 \pi$ (Fig.
\ref{fig5}(c)). The corresponding plot of $\mathcal D$ for $l=1$ and
$l=2$ as a function of $n$ (Fig.\ \ref{fig5}(a)) shows two different
relaxation behavior in accordance with our analysis. The presence of
such line of zeroes excludes the possibility of re-entrant behavior
since an entire line of zeroes can not generically vanish at a
single $\omega$ as it is varied.

To understand the point discussed above in a bit more general
setting, let us consider a class of Hamiltonians which has the form
\begin{eqnarray}
H_{\vec k} &=& h[g_1(k_x) + \alpha_1 g_0(k_y), g_2(k_x)+ \alpha_2
g_2(k_y) ...; \beta(t)] \nonumber\\
&& \equiv h[g_p (k_x)+ \alpha_p g_p(k_y);\beta(t)]
\end{eqnarray}
for $1\le p\le p_{\rm max}$, where $g_p$ are arbitrary functions of
$k_x$ or $k_y$ and $\alpha_p$ are parameters of $H_{\vec k}$. For
example, for the Kitaev model $p_{\rm max}=2$, $g_1 = \cos(k_i)$,
$g_2=\sin(k_i)$, $\beta(t)= J_3(t)/J_1$, and $ \alpha_1=\alpha_2
=J_2/J_1$. It is easy to see that the drive does not change this
functional form; thus $U_{\vec k}$ and hence $H_{\vec k F}$ and
$|\epsilon_{\vec k}|$ retains the same structure
\begin{eqnarray}
|\epsilon_{\vec k}| = h'[g_p (k_x)+ \alpha_p g_p(k_y);T].
\end{eqnarray}
Such a functional form guarantees that if $\partial |\vec
\epsilon_{\vec k}|/\partial k_x =0$ so is $\partial |\vec
\epsilon_{\vec k}|/\partial k_y$. This implies that the zeroes of
$\partial |\vec \epsilon_{\vec k}|/\partial k_x$ (or equivalently
$\partial |\vec \epsilon_{\vec k}|/\partial k_y$) forms a 1D curve
in the 2D Brillouin zone leading to a line of zeroes of $\nabla
|\vec{\epsilon}_{\vec{k}}|$. The generalization of this result for
$d>2$ may lead to $d' < d$ hypersurfaces of zeroes of $\nabla
|\vec{\epsilon}_{\vec{k}}|$ in a $d$-dimensional Brillouin zone.

\subsection{Phase diagram for the Ising model}
\label{dyntrb}

In this section, we sketch the phase diagram for dynamical phases
for the 1D transverse field Ising model as a function of the initial
transverse field $g_i$ and the drive frequency $\omega$ for the
square pulse protocol. In what
follows, we shall either use this protocol or the
periodic kick protocol for which
\begin{eqnarray}
g(t)=g_{0}+g_{1}\sum_{n=0}^{\infty}\delta(t-nT). \label{delk}
\end{eqnarray}
where the analysis is simplified further.

\begin{figure}
{\includegraphics[width=\hsize]{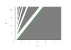}}
 \caption{ Phase diagram for the dynamical phases for the
square pulse drive protocol where $g_f=0$. The grey (white) regions
correspond to a relaxational behavior of $n^{-3/2}$ ($n^{-1/2}$)
 of local quantities to their corresponding steady state values.
The green dots have been obtained using Eq.~\ref{condtrans}.
\label{fig6a}}
\end{figure}

\begin{figure}
{\includegraphics[width=\hsize]{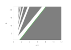}}
 \caption{ Phase diagram for the dynamical phases for the
square pulse drive protocol where $g_f=2$. The color scheme is the same
as in Fig.~\ref{fig6a}.
The green dots have been obtained using Eq.~\ref{condtrans}. Note that the 
equilibrium
critical point at $g=1$ is never crossed during the dynamics here.
\label{fig6b}}
\end{figure}

Such a phase diagram for the square pulse drive
for $g_f=0$ is shown in Fig.\ \ref{fig6a}
and for $g_f=2$ in Fig.\ \ref{fig6b}.
Note that in the latter case, the equilibrium critical point at $g=1$
is never crossed during the dynamics.
In both cases, we see intermittent regions of $n^{-3/2}$ (grey) and $n^{-1/2}$
(white) relaxation to the final steady state.
The relaxation is always $n^{-3/2}$ in
the $\omega \rightarrow \infty$ limit and $n^{-1/2}$ in the
$\omega \rightarrow 0$ limit inspite of multiple re-entrances present
in the phase diagram, consistent with the general
argument presented in Sec.~\ref{1dstuff}.
Furthermore, the number of re-entrant regions increase
as the amplitude of the drive $g_i$ is increased in both cases.
We also show the perfect match for the location of the first
dynamical phase transition
as the frequency is reduced at any fixed $g_i$ using Eq.~\ref{condtrans}
in the phase diagrams
presented in Fig.~\ref{fig6a} and Fig.~\ref{fig6b}.

\begin{figure}
{\includegraphics[width=\hsize]{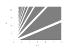}}
 \caption{ Phase
diagram for the delta function kick protocol where we fix
$g_{1}=1$. The color scheme is the same
as in Fig.~\ref{fig6a} and Fig~.\ref{fig6b}.
\label{fig7}}
\end{figure}

A similar plot for the delta function kicked
protocol given by Eq.\ \ref{delk} is shown in Fig.\ \ref{fig7}.
Next, we analyze the phase diagram obtained by using the delta
function kick protocol analytically. The evolution operators for
this protocol is given by
\begin{eqnarray}
U_k(T,0) &=& e^{- i g_1 \tau_3} e^{-iT((g_0-\cos(k))\tau_3
+\sin(k) \tau_1)} \nonumber\\
&=&  \left( \begin{array}{cc} \alpha_k & -\beta_k^{\ast} \\
\beta_k & \alpha_k^{\ast} \end{array} \right) \nonumber\\
\alpha_k &=& e^{-ig_{1}}(\cos(\Phi_k)-i\sin(\Phi_k)\hat{n}_{k z})
\nonumber\\
\beta_k &=& -ie^{-ig_{1}}\hat{n}_{k x}\sin(\Phi_k) \label{evolu1}
\end{eqnarray}
where we have $\epsilon_k=\sqrt{(g_{0}-\cos(k))^{2}+(\sin(k))^{2}}$,
$\hat{n}_{kx}=\sin(k)/\epsilon_k$, $\hat{n}_{ky}=0$,
$\hat{n}_{kz}=(g_{0}-\cos(k))/\epsilon_k$, and $\Phi_k=T
\epsilon_k$. Using Eq.\ \ref{evolu1}, one can find the Floquet
spectrum to be
\begin{eqnarray}
\alpha_{kF}=\frac{1}{T}\arccos[{\cos(\Phi_{k}+g_{1})+(1-\hat{n}_{kz})\sin(\Phi_{k})}]\label{eq:9}
\end{eqnarray}
For further analysis, we note that $\hat{n}_{kz} \to 1$ for
$g_{0} \gg 1$ allowing a perturbative expansion of $\alpha_{k F}$. This leads to
\begin{eqnarray}
\label{eq:A}
\alpha_{kF}=\epsilon_{k}+\frac{g_{1}}{T}-\frac{\sin^{2}k\sin(\Phi_{k})\sin(g_{1})}{2T(g_{0}-\cos(k))^{2}|\sin(\Phi_{k}+g_{1})|}
\end{eqnarray}

\begin{figure}
{\includegraphics[width=\hsize]{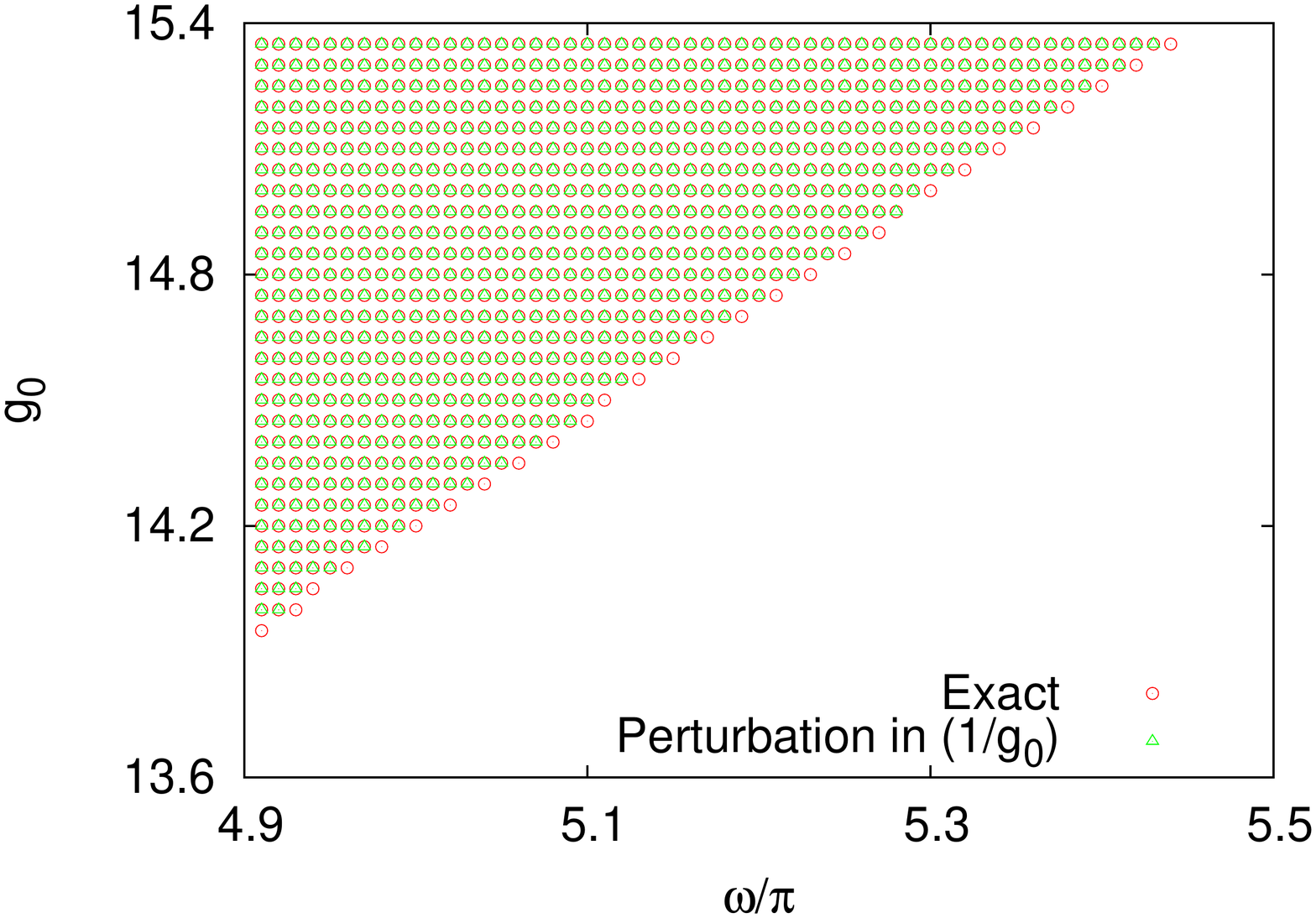}}
 \caption{
Phase diagram for the delta function kick protocol where we fix
$g_{1}=1.0$ and vary $g_{0}$. The white region indicates $n^{-3/2}$
relaxation of local quantities to their final steady state values.
The green points denote the perturbative results obtained from
solution of Eq.\ \ref{eq:10} for large $g_0$ while the red points
denotes the exact results for the $n^{-1/2}$ relaxation behavior.
\label{fig8}}
\end{figure}

Having obtained $\alpha_{kF}$, we now look for zeroes of $
d(\alpha_{kF})/dk=0 $. If these occur only at $k=0,\,\pi$, the
relaxation behavior of the system will demonstrate $n^{-3/2}$
behavior. Thus $n^{-1/2}$ relaxation occurs when the zeroes of $d
(\alpha_{k F})/dk$ occurs at other values of $k$; the condition for
this can be shown, after some straightforward algebra, to be
\begin{eqnarray}
\frac{g_0} {\epsilon_k} \left(1-\frac{\sin^{2}g_{1}\sin^{2}k\text{
Sgn}(\sin(\Phi_k+g_1))}{2g_{0}^{2}\sin^{2}(\Phi_k+g_1)} \right)
\nonumber\\=\frac{\cos{k}
\sin{\Phi_{k}}\sin{g_{1}}}{g_{0}^{2}T|\sin{(\Phi_{k}+g_{1}})|}
\label{eq:10}
\end{eqnarray}
where Sgn denotes the signum function.
We note that for large $g_0$, Eq.\ \ref{eq:10} is satisfied within a
small parameter range around the point $ (\Phi_{k}+g_{1})= m
\pi$ (where $m \in Z$) for which $g_0 |\sin(\Phi_k + g_1)| \sim
1$. This implies that the density of re-entrant regions scale as $g_0$
at large $g_0$,  which explains the increase in the number of 
re-entrant regions as the drive amplitude is increased. 
A computation of the phase diagram with exact numerical result
is shown in Fig.\ \ref{fig8}; we find the perturbative and numerical
results match well in the large $g_0$ regime, and thus $1/g_0$ acts as
a suitable expansion parameter unlike $1/\omega$ to get the dynamical
transitions.

\section{Entanglement in the steady state}
\label{ensteady}

In this section, we shall address the entanglement entropy in
the final steady state and calculate $S_{\infty}(l)$.
In the $n \rightarrow \infty$ limit, the
calculation becomes simple since the system can be described
locally by the Diagonal ensemble. 
The computation proceeds in the following manner.
Given an initial state at $t=0$,
$|\psi \rangle = \otimes_{\vec k}|\psi_{\vec k}(t=0) \rangle$, we
can express each $|\psi_{\vec k}(t=0) \rangle$ in terms of the
eigenvectors $|1_{\vec{k}} \rangle, |2_{\vec k} \rangle$ of the
Floquet Hamiltonian $H_{\vec{k} F}$. Then, it follows that
\be
\langle \psi_{\vec{k}}(nT)| O_{\vec{k}}|\psi_{\vec{k}}(nT) \rangle =
|a_{1\vec{k}}|^2 \langle 1_{\vec{k}}|O_{\vec{k}}|1_{\vec{k}} \rangle + \nonumber \\
|a_{2\vec{k}}|^2 \langle 2_{\vec{k}}|O_{\vec{k}}|2_{\vec{k}} \rangle
+a_{1\vec{k}}^* a_{2\vec{k}} e^{-i2n|\vec{\epsilon}_{\vec{k}}|T} \langle 1_{\vec{k}}|O_{\vec{k}}|2_{\vec{k}} \rangle  + \nonumber \\
a_{1\vec{k}} a_{2\vec{k}}^* e^{i2n|\vec{\epsilon}_{\vec{k}}|T} \langle 2_{\vec{k}}|O_{\vec{k}}|1_{\vec{k}} \rangle
\label{DEsteps}
\ee
where $a_{1(2)\vec{k}} = \langle 1(2)_{\vec k} |
\psi_{\vec k}(t=0) \rangle$.  It is then justified to drop
the rapidly oscillating cross-terms when $n \rightarrow \infty$
for calculating the expectation
value of any local (in space) operator in the thermodynamic limit.
The steps are explicitly shown for the correlation
matrix $\mathcal{C}_\infty (l)$ in Appendix \ref{secmeasure}.
Dropping the cross-terms in Eq.~\ref{DEsteps} leads to a mixed density matrix
in the orthonormal basis $|1(2)_{\vec{k}} \rangle$ which is called the Diagonal
Ensemble and the $n \rightarrow \infty$ results for all local quantities and
the entanglement entropy of finite subsystems shown here have been obtained
from the same.

\begin{figure}
{\includegraphics[width=\hsize]{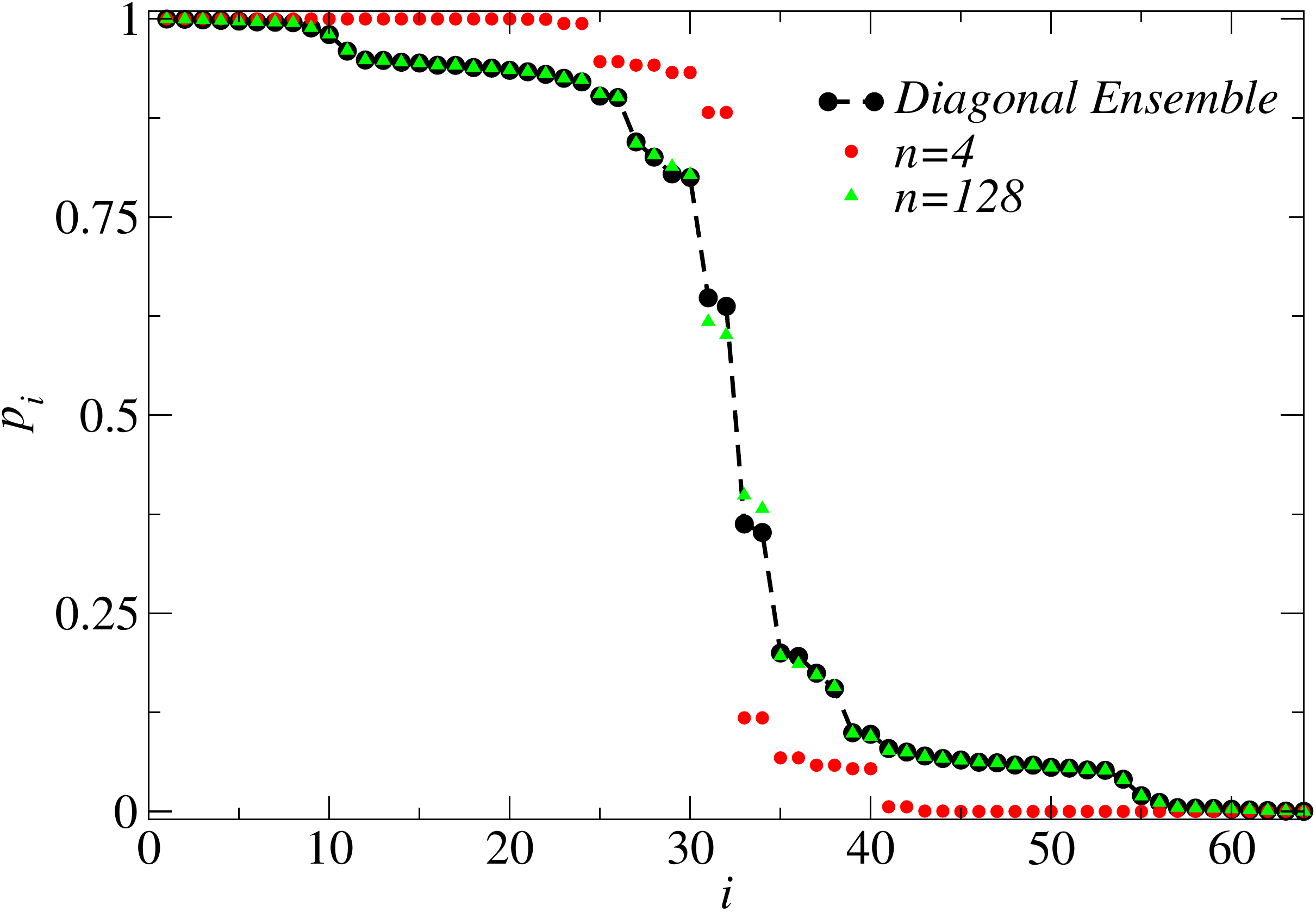}}
\caption{The approach of the eigenvalues of the correlation matrix of
a subsystem of size $l=32$ to the Diagonal ensemble results as a function of $n$
for the 1D Ising model driven according to a square pulse protocol
with $g_i=3$, $g_f=2$ and $T=2$. At small
$n$, the majority of $p_i$ are either $0$ or $1$.
 \label{fig9}}
\end{figure}

The entropy of the system in the Diagonal Ensemble, which we denote
by $S_{\mathrm{DE}}$, may be
readily calculated.
Dropping the cross-terms from Eq.~\ref{DEsteps} in the thermodynamic
limit, we obtain
 \be
\frac{S_{\mathrm{DE}}}{L^d} &=& s_{\mathrm{DE}}=\frac{2}{(2\pi)^d} \int_{\vec{k} \in BZ/2} d^dk s_{\vec{k}} \nonumber \\
s_{\vec{k}}&=& -p_{\vec{k}}\log p_{\vec{k}}-(1-p_{\vec{k}})\log(1-p_{\vec{k}})
\label{DE_ent}
\ee
where $p_{\vec{k}} = |a_{1\vec{k}}|^2$ and $s_{\vec{k}}$ is the
entropy at momentum $\vec{k}$ (which is a good quantum number here since
we look at the entropy of the full system in Eq.~\ref{DE_ent}).
$s_{\vec{k}}$ equals zero if $p_{\vec{k}}=0,1$ and is
maximized to $\ln(2)$ for $p_{\vec{k}}=1/2$. If the integral
$\int_{\vec{k} \in BZ/2} d^dk s_{\vec{k}}$ is
finite, then the entropy $S_{\mathrm{DE}}$ clearly scales extensively.
We see that the eigenvalues of the matrix $\mathcal{C}_n(l)$
approach those of $\mathcal{C}_{\infty}$ for any $l \ll L$ when $n$ is
large enough (Fig.~\ref{fig9}). Thus, the entanglement entropy density
$S_{\infty}(l)/l^d$ when $l \rightarrow \infty$ and $l/L \rightarrow 0$
coincides
with $s_{\mathrm{DE}}$ (Eq.~\ref{DE_ent}).

\begin{figure}
{\includegraphics[width=\hsize]{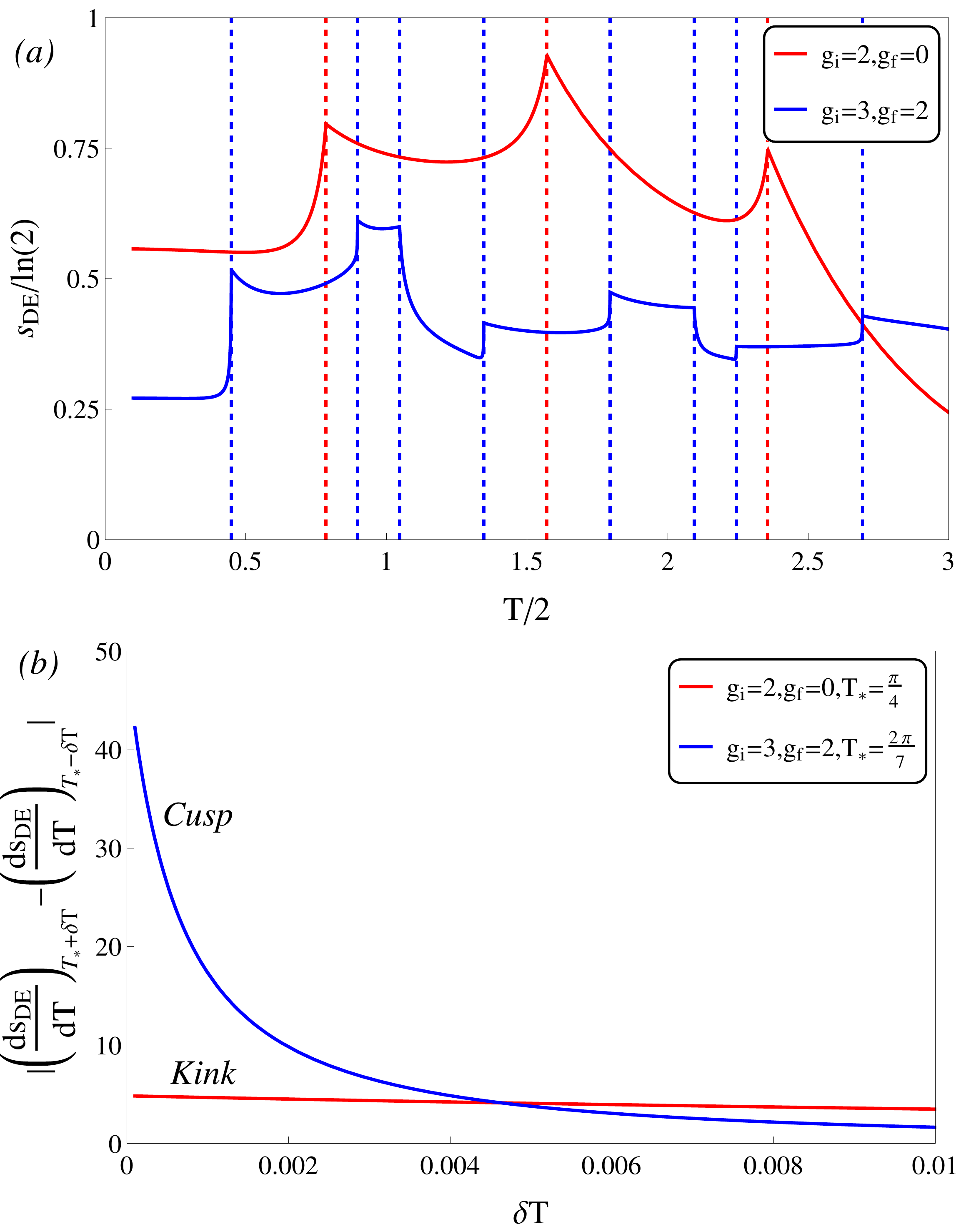}} \caption{(a) The entropy
per site in the diagonal ensemble $s_{\mathrm{DE}}$ as a function of
$T=2\pi/\omega$ for the 1D transverse field Ising model. The initial
state is taken to be $\sigma_i^x=1$ for all $i$, and the system is
then driven using a square pulse protocol with $g_i=2, g_f=0$ (red
curve) and $g_i=4, g_f=2$ (blue curve with $s_{\mathrm{DE}}$ multiplied 
by a factor of two in this case). The locations of the
singularities (dashed vertical lines, where red lines are for
$g_i=2,g_f=0$ and blue lines are for $g_i=3,g_f=2$) have been
obtained using Eq.~\ref{FBcross}. (b) Derivative singularities which
are either cusps or kinks exist in $s_{\mathrm{DE}}$ at the driving
time periods $T_\ast$ where the Floquet bands cross each other.
 \label{fig10}}
\end{figure}

The entropy per site $s_{\mathrm{DE}}$ shows a  non-monotonic
behavior as a function of $\omega$. More interestingly, there are
sharp features at particular values of $\omega$ 
in $s_{\mathrm{DE}}$ (Fig.~\ref{fig10}(a)) in
$d=1$. We further show here that these features are in fact
derivative singularities that are either cusps or kinks
(Fig.~\ref{fig10}(b)). The analysis is greatly simplified if the
initial state $|\psi_{k} (t=0)\rangle$ is taken to be $(0,1)^T$ at
each $k$. For the 1D Ising model, this corresponds to the ground
state at $g \rightarrow \infty$. Then, it is straightforward to show
that \be s_{k} = -\frac{1-\hat{n}_{k3}}{2} \ln
\left(\frac{1-\hat{n}_{k3}}{2} \right)- \frac{1+\hat{n}_{k3}}{2} \ln
\left(\frac{1+\hat{n}_{k3}}{2} \right) \label{formula_s} \ee Thus,
$s_{k}=0$ if $\hat{n}_{k3} = \pm 1$ and attains its maximum value of
$\ln(2)$ when $\hat{n}_{k3}=0$.

\begin{figure}
{\includegraphics[width=\hsize]{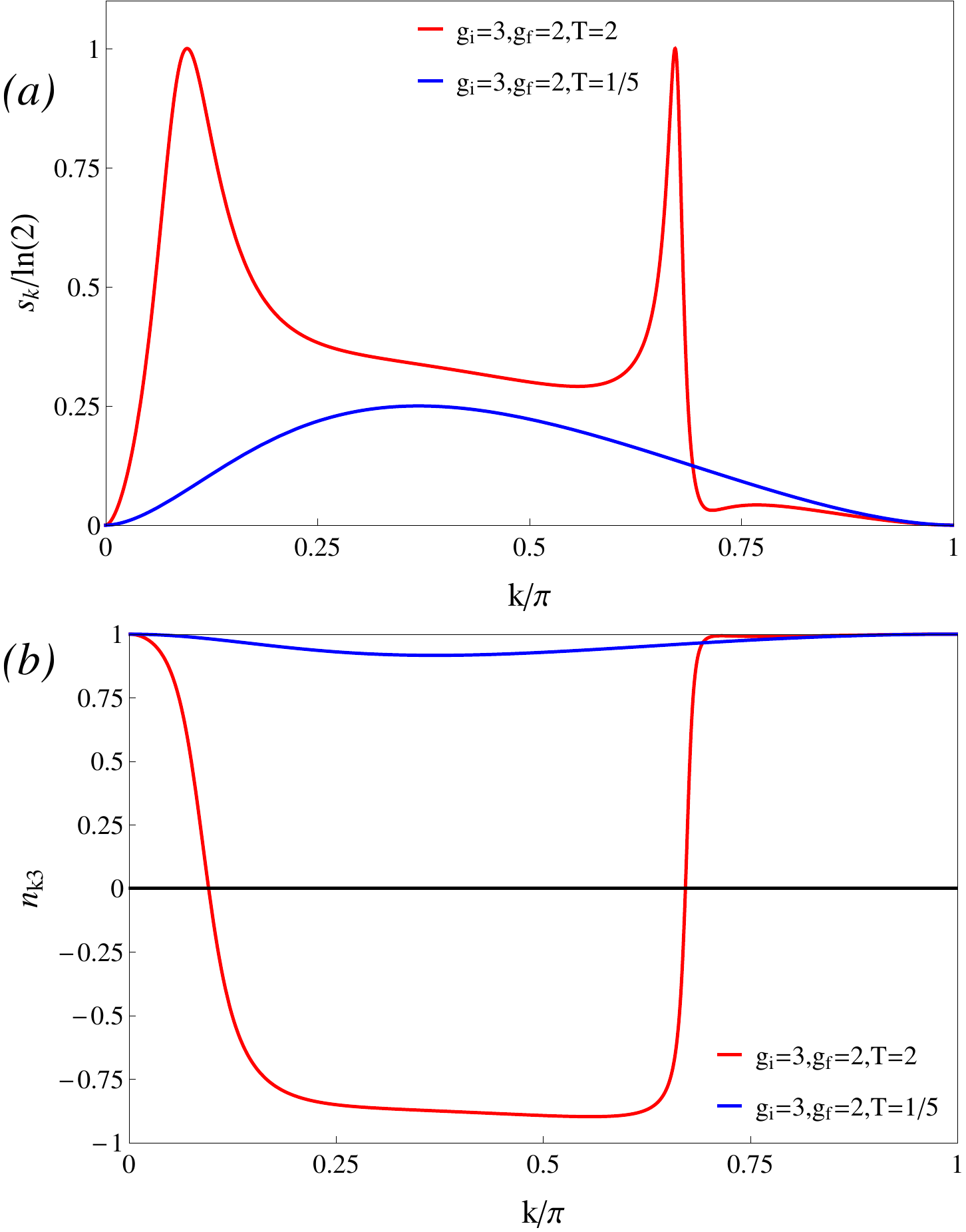}}
 \caption{ (a) $s_{k}$ as a function of $k$ for the square pulse
protocol with $g_i=3$, $g_f=2$ and drive time period
$T=1/5$ (blue curve) and $T=2$ (red curve). 
(b) $s_{k}$ attains its maximum value of $\ln(2)$ whenever $\hat{n}_{k3}=0$. 
\label{fig11}}
\end{figure}

In the large $\omega$ limit,
the Floquet Hamiltonian $\mathcal{H}_{kF}$
can be well approximated by the average Hamiltonian over one cycle and thus
$\hat{n}_{k3}$ does not have any zeroes in $k \in [0,\pi]$
except at $k=0$ or $k=\pi$ in special cases when $g_{\mathrm{av}} =
(1/T)\int_0^T g(t)dt
=\pm 1$.
As $\omega$ is decreased, new zeroes of $\hat{n}_{k3}$
are generated between the zone
boundaries $k=0$ and $k=\pi$ (Fig.~\ref{fig11}(b))
which leads to the presence of newer peaks in
the function $s_{k}$ (Fig.~\ref{fig11}(a)). 
These zeros occur at $k=k_0$ such that
\begin{eqnarray}
\cos(\theta_{k=k_0}) &=& 0 \quad {\rm or} \quad
\sin(\alpha_{k=k_0})=0. \label{zeroe3cond}
\end{eqnarray}
The number of zeroes in the function $n_{k3}$ scales as $1/\omega$ when
$\omega \rightarrow \infty$. In $d>1$, the new zeroes in $\hat{n}_{\vec{k}3}$
appear along $(d-1)$-dimensional hypersurfaces in the $d$-dimensional BZ.

\begin{figure}
{\includegraphics[width=\hsize]{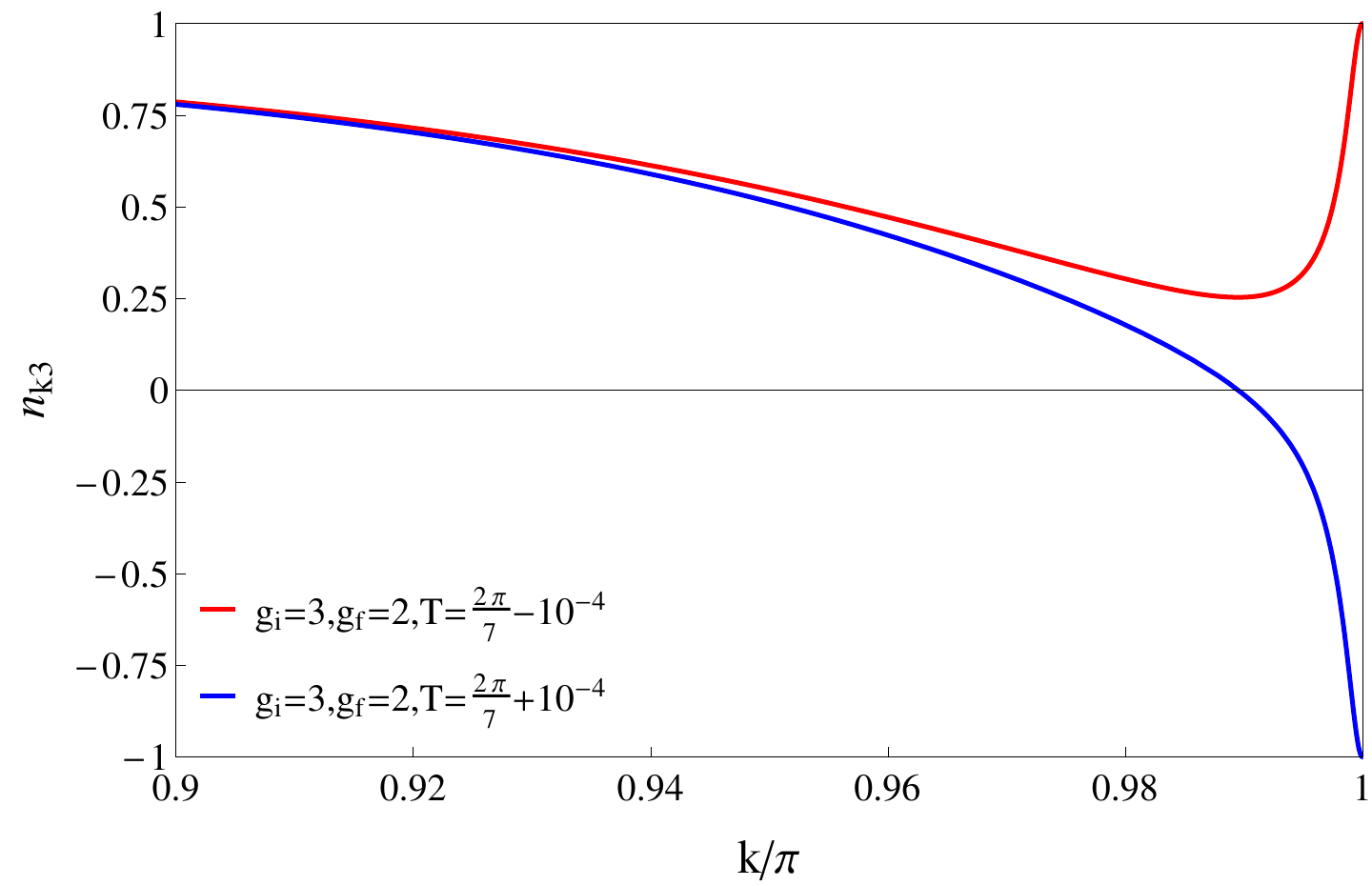}} \caption{The behavior of
$\hat{n}_{k3}$ when $k \rightarrow \pi$ and $T = T_{\ast}
\pm 10^{-4}$ where $T_{\ast}=2\pi/\omega_{\ast}$ corresponds to a Floquet band
crossing at $k=\pi$. The drive follows a square pulse protocol with $g_i=3$ 
and $g_f=2$. \label{fig12}}
\end{figure}

We now consider the total number of zeroes in $\hat{n}_{k3}$ where
$k \in (0,\pi)$ (thus excluding $k=0$ and $k=\pi$). The frequencies
$\omega_{\ast}=\frac{2\pi}{T_{\ast}}$ across which this number
changes by one are precisely the locations of the singularities in
$s_{\mathrm{DE}}$ in 1D (dashed lines marked in
Fig.~\ref{fig10}(a)). The new zero in $\hat{n}_{k3}$ has to
enter/exit from either of the zone boundaries at $k=0$  or $k=\pi$.
However, $\hat{n}_{k3}$ equals $\pm 1$ at the boundaries and hence,
the only way a zero can enter/exit is if $\hat{n}_{k3} = \pm 1$
jumps to $-\hat{n}_{k3} = \mp 1$ at the zone boundary from where the
new zero enters/exits. Hence, the function $\hat{n}_{k3}$ is
necessarily singular at $\omega = \omega_{\ast}$ with \be
\left(\frac{d\hat{n}_{k3}}{d\omega}\right)_{\omega \rightarrow
\omega_{\ast}^-,k \rightarrow 0/\pi} \neq
\left(\frac{d\hat{n}_{k3}}{d\omega}\right)_{\omega \rightarrow
\omega_{\ast}^+,k \rightarrow 0/\pi}. \ee The behavior of
$\hat{n}_{k3}$ as a function of $k$ is shown in Fig.~\ref{fig12} in
the vicinity of one such $\omega_{\ast}$. The singularity in
$\hat{n}_{k3}$ leads to the resulting derivative singularity in
$s_{\mathrm{DE}}$ through the relation: \be \left(\frac{d
s_{\mathrm{DE}}}{d\omega}\right)  = \frac{1}{2}\int_0^{\pi} dk \log
\left(\frac{1-\hat{n}_{k3}}{1+\hat{n}_{k3}} \right) \frac{d
\hat{n}_{k3}}{d \omega}, \ee which we show can be either a cusp or a kink
singularity in Fig.~\ref{fig10}(b).

We now show that these $\omega_{\ast}$ are located at the
frequencies where the Floquet bands $\pm \Phi_{k}$, defined by
$U_{k}=\exp(\pm i \Phi_{k})$, cross each other. This is equivalent
to the condition that $U_{k} = \pm I$. For the class of models we
consider here (Eq.~\ref{hamdef1s}), it was shown in Ref.
\onlinecite{dsenks1} that such crossings can only occur at those
momenta $k_0$ where $\Delta_{k_0}=0$, and thus $U_{k_0}$ and
$\hat{n}_{k_{0}3}$ may be straightforwardly calculated for any
protocol. Specializing to 1D, we use Eq.~\ref{ueq3a} to obtain
$\hat{n}_{k3}$ at $k=0,\pi$ for the square pulse protocol which
gives \be
\hat{n}_{k=0,3} &=& \mathrm{Sgn} \left(\sin (T(g_{\mathrm{av}}-1))\right) \nonumber \\
\hat{n}_{k=\pi,3} &=& \mathrm{Sgn} \left(\sin (T(g_{\mathrm{av}}+1))\right)
\ee
 Thus, $\hat{n}_{k=0/\pi,3}$ jumps from $\pm 1$ to $\mp 1$ whenever
\be
\omega_{\ast} = \frac{2(g_{\mathrm{av}} \pm 1)}{n}
\label{FBcross}
\ee
where $n$ is any positive integer.
This is also precisely the condition to obtain $U_{k}=\pm I$ at $k=0/\pi$, and
hence the singularities in $s_{\mathrm{DE}}$ occur at the Floquet band
crossings (as shown in Fig.~\ref{fig10}(a)). Since the only details of the
drive protocol that enters $U_k$ when $\Delta_{k_0}=0$ are $g_{\mathrm{av}}$, $T$
and $k_0$, the condition for $\omega_{\ast}$ (Eq.~\ref{FBcross}) stays
unchanged for any drive protocol in 1D. These arguments can be easily
carried over to $d>1$ and the Floquet band crossings may again lead to
singularities in $S_{\mathrm{DE}}/L^d$, though such singularities
will get weaker (if they survive at all) because
of the $d$-dimensional integrals in $\vec{k}$ space.

\section{Discussion}
\label{diss}

In this work, we have studied entanglement generation and pointed
out the presence of a dynamical transition in a class of
periodically driven integrable models. Our work shows that the
states of such systems crosses over from an area- to volume-law
entanglement entropy in a complex manner depending on the drive
frequency and the number of drive cycles. This leads to the presence
of non-area and non-volume like behavior of $S(l)$: $S(l) \sim
l^{\alpha}$ with $d-1 \le \alpha \le d$; we note that our work shows
that states with such non-area and non-volume law entanglement can
be generated in a controlled manner by tuning the drive frequency.

We also study the relaxation behavior of the correlation function
and the density matrices of such periodically driven systems to
their steady state (GGE) values as a function of the number of drive
frequency. In doing so, we unravel a frequency controlled dynamic
transition between two phases of the system. In the high frequency
regime, $S_n$ decays to $S_{\infty}$ as $ n^{-d/2-1}$ while below a
critical frequency, it does so as $n^{-d/2}$. We show that this
transition can be understood as a change in topology of the Flouqet
spectrum and leads to change in behavior of local correlation
function. This is in sharp contrast to other class of dynamic
transition studied in the literature \cite{pol1,sub1, sub2,
sangita1} which is associated with non-analyticities (also known as
Fischer zeroes) of the dynamical free energy of the system $f(z) = -
\lim_{L \to \infty} \ln(F(z))/L^d$, where $z$ is obtained by
analytic continuation of time $t$ in the complex plane and which do
not leave their mark on local correlation function. We present a
detailed phase diagram corresponding to these transition and provide
an analytical expression for one of the phase boundaries for
the square pulse protocol which matches exact numerics. We also show 
that a perturbative expansion in terms of the inverse of the 
drive amplitude captures 
the dynamical transitions reliably in the large amplitude limit, whereas
an expansion in $1/\omega$ fails to do so. 

Finally, we also study the entanglement entropy of such driven
systems in the steady state described by the diagonal ensemble. We
find that in 1D $S_{\infty}/L$ displays singular cusp/kink like features as
a function of $\omega$ at special values of the drive frequencies:
$\omega= \omega^{\ast}$. We identify the presence of such cusp/kink
like features with Floquet band crossings and provide an explicit
expression of $\omega^{\ast}$ for arbitrary drive protocol. We note
that for a large number of drive protocols such Floquet band
crossings indicate topological transitions of driven systems; thus
our results indicate that the steady state entanglement entropy
bears signatures of such topological transitions. We also note that
our arguments in this regard may be carried over to integrable
systems with $d>1$. However, it is not clear whether such singular
features in $S_{\infty}/L^d$ will survive due to effect of higher
dimensional momentum integrals involved in computation of
$S_{\infty}/L^d$.

We note that our work can be verified using simple experiments.
Recently second R\'enyi entropy $S^{(2)} = -{\rm Tr}[\rho^2]$ has
been measured via measurement of overlap of two quantum many-body
states of ultracold atom systems \cite{rajexp}. We propose analogous
measurement of $S^{(2)}_n$ (which has similar properties as $S_n$)
as function of $n$ and predict its $(\omega/n)^{(d+2)/2}
[(\omega/n)^{d/2}]$ scaling for fast [slow] drives.  The different
dynamical regimes and their re-entrant behavior may also be
experimentally observed via magnetization ($\langle \sigma_x
\rangle_n$) measurement of the periodically driven Ising model. We
note here that such ultracold atom setups are currently
experimentally feasible \cite{rev7,ess1}.

In conclusion, we have demonstrated controlled realization of
quantum states with non-area and non-volume law entanglement entropy
and predicted the presence of two distinct dynamical phases
separated by transitions arising from change in topology of the
system's Floquet spectrum. We have also demonstrated that the steady
state entanglement entropy, for 1D integrable systems, displays
singular features at special frequencies which is associated with
Floquet band crossings. We have suggested specific experiments which
can test our theory. 

Finally, our work leads to several
possibilities of future extensions such as study of possible
realization of such dynamical phases in non-integrable and models
with long-range interactions.

\appendix

\section{ Ising and Kitaev model} \label{iskit}

In this section, we sketch the connection of Eq.\ \ref{hamdef1s}
 in the main text
with a large class of spin models. For example in $d=1$, the
transverse field Ising model has the Hamiltonian \be H_{\rm Ising}
=-\sum_{j=1}^N (h \sigma_j^x + \sigma_j^z \sigma_{j+1}^z),
\label{tfim} \ee where $\sigma^{x,y,z}$ are the usual Pauli
operators, $h$ denotes the transverse field, and we have scaled all
quantities by the nearest neighbor interaction $J$. The ground state
of this model is ferromagnetic when $-1<h<1$ and paramagnetic
otherwise. There are thus two critical points in this model at $h =
\pm 1$. It turns out that $H_{\rm Ising}$ allows a fermionic
representation in terms of $H$ with $b_k=\cos(k)$, $\Delta_k =
\sin(k)$, $g=h$, and $\psi_k = (c_k, c_{-k}^{\dagger})^T$ via a
Jordan-Wigner representation given by \be
\sigma_n^x &=& 1-2c^{\dagger}_n c_n \nonumber \\
\sigma_n^z &=& -(c_n+c_n^{\dagger})\prod_{m<n}(1-2c^{\dagger}_m
c_m), \label{jw} \ee where $c_n$ is the Fermionic annihilation
operator on site $n$ and $c_k$ denotes its Fourier transform
\cite{subir1}. This fermionic representation of $H_{\rm Ising}$ has
been used, with $g(t)$ following a square pulse protocol, to
generate the results shown in Fig.\ \ref{fig1}, Fig.\ \ref{fig2}, 
Fig.\ \ref{fig3} and Fig.\ \ref{fig4} in the main text.

\begin{figure}
{\includegraphics[width=\hsize]{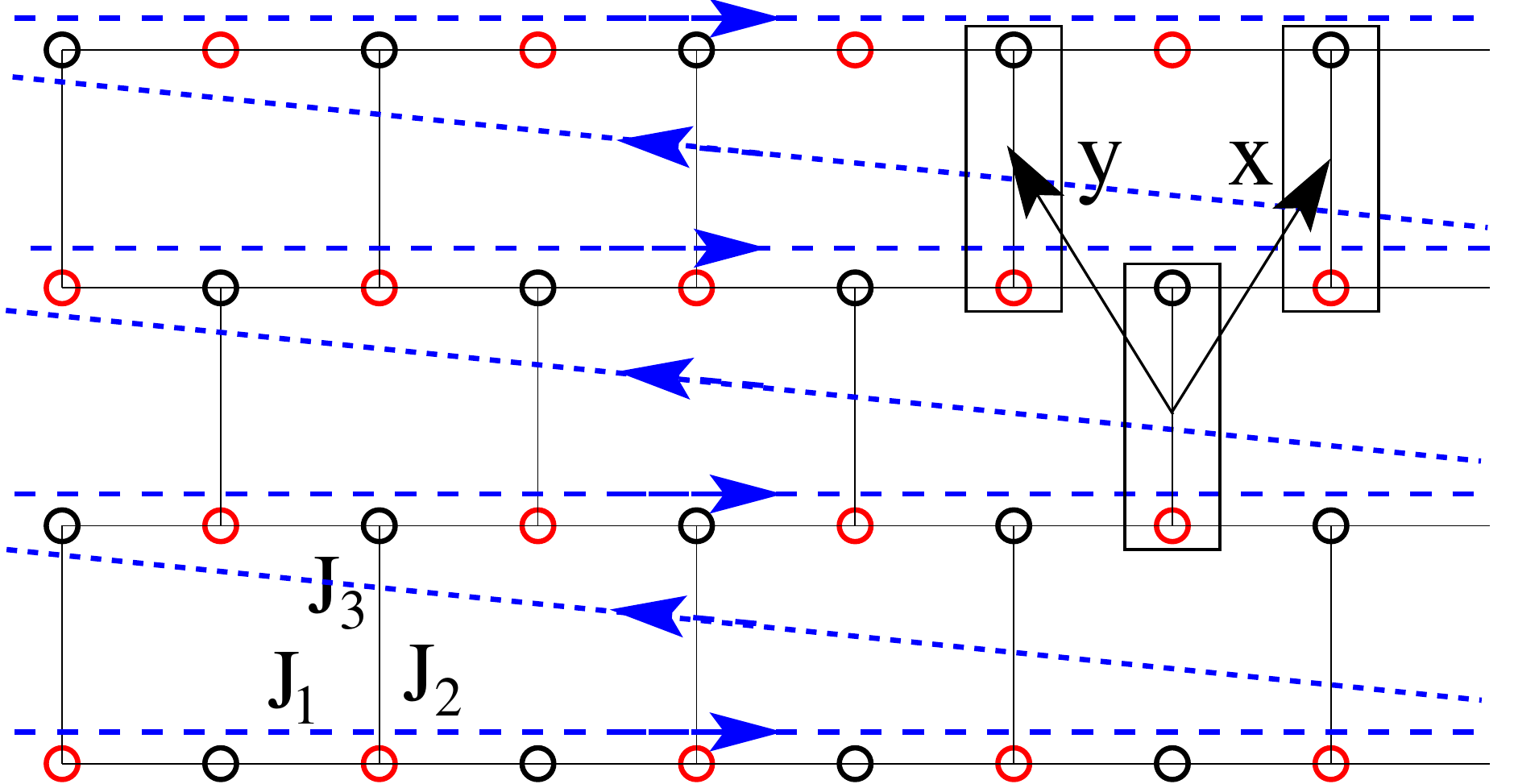}} \caption{Brick-wall
lattice which is equivalent to the 2D hexagonal lattice and the
three types of interactions in the Kitaev model. Also shown is the
contour for the Jordan-Wigner transformation that is employed for
the fermionization of the model.
\label{supp1}}
\end{figure}

A similar correspondence can be obtained in $d=2$ by considering the
Kitaev model \cite{kitaev1,feng1,nussinov1,sengupta1} whose
Hamiltonian is given by \be H_{\rm 2D} &=& \sum_{j+l={\rm even}} ~(~
J_1 \si_{j,l}^x \si_{j+1,l}^x ~+~
J_2 \si_{j-1,l}^y \si_{j,l}^y \non \\
& & ~~~~~~~~~~~~~~+ J_3 \si_{j,l}^z \si_{j,l+1}^z ~).
\label{kham1}\ee This Hamiltonian describes a spin model on a
hexagonal 2D lattice, where $j$ and $l$ denote the column and row
indices of the brick-wall lattice which is an alternative
representation of the hexagonal lattice (Fig.~\ref{supp1}). The
Kitaev model can be fermionized in an analogous manner to the Ising
model by taking a Jordan-Wigner transformation along a
one-dimensional contour that threads the entire lattice and passes
through each lattice site exactly once (Fig.~\ref{supp1}).
Introducing a pair of Majorana fermions for each fermion, the Kitaev
model then reduces to a model of Majorana fermions coupled to Z$_2$
gauge fields. The crucial point that makes the solution of Kitaev
model feasible is that the Z$_2$ fields, which we denote by
$\alpha_r$, commute with $H_{\rm 2D}$, so that all the eigenstates
of $H_{\rm 2D}$ can be labeled by their specific values
($\alpha_r=\pm 1$). It has been shown that for any value of the
parameters $J_i$, the ground state of the model always corresponds
to all $\alpha_r$ equal to $1$. Since $\alpha_r$ is a constant of
motion, the dynamics of the model starting from any ground state
never takes the system outside the manifold of states with
$\alpha_r=1$. The Majorana fermions can then be combined pairwise on
each $J_3$ bond (as shown in Fig.~\ref{supp1}) to give an equivalent
free fermion Hamiltonion on the square lattice~\cite{nussinov1}: \be
H_{\mathrm 2D} &=& J_1 \sum_r (c^{\dagger}_r+c_r)(c^{\dagger}_{r+\hat{x}}-c_{r+\hat{x}}) \nonumber \\
&+&J_2 \sum_r (c^{\dagger}_r+c_r)(c^{\dagger}_{r+\hat{y}}-c_{r+\hat{y}}) \nonumber \\
&+& J_3 \sum_r \alpha_r (2c^{\dagger}_rc_r-1) \ee where $\alpha_r=1$
if the initial state is a ground state. Thus, like in the 1D case,
$H_{\rm 2D}$ allows a fermionic representation in terms of $H$ with
$b_{\vec{k}}=J_1\cos(k_x)+J_2\cos(k_y)$, $\Delta_{\vec{k}} =
J_1\sin(k_x)+J_2\sin(k_y)$, $g=J_3$, and $\psi_{\vec{k}} =
(c_{\vec{k}}, c_{-\vec{k}}^{\dagger})^T$. Note that the static Z$_2$
gauge fields give an additional additive contribution to the
entanglement entropy that follows the area law \cite{yao1} and does
not change under the dynamics, whereby we can ignore this (static)
contribution while considering the generation of entanglement in a
periodic drive.

The energy spectrum of $H_{\rm 2D}$ consists of two bands with
energies
\be E_{\vk}^\pm &=& \pm ~2 ~[(J_1 \sin (k_x) + J_2 \sin (k_y))^2 \non \\
&& ~~+ (J_3 - J_1 \cos (k_x) + J_2 \cos (k_y)^2 ]^{1/2} .
\label{hk1} \ee We note for $|J_1-J_2|\le J_3 \le (J_1+J_2)$, these
bands touch each other so that the energy gap $E_{\vk}^+ -
E_{\vk}^-$ vanishes for special values of $\vk$ leading to the
gapless phase of the model. This fermionic representation of $H_{\rm
2D}$ has been used, with $g(t)$ following a square pulse protocol,
to generate the results shown in Fig.\ \ref{fig5} in the main text.

Before ending this section, we note that it is possible to obtain an
analytic solution for $|\psi_k (nT)\rangle$ for the square pulse
protocol as follows. Within each cycle, for a given $g_i$, $g_f$ and
$T$, one constructs a unitary matrix matrices $U(\vec{k},g_i,g_f)$
which evolve an arbitrary initial $|\psi_{k}(0) \rangle$ to the
state after time $T$ by \be |\psi_{\vec k} (T)\rangle  = U({\vec
k},g_i,g_f) |\psi_{\vec k}(0) \rangle \ee Thus, after $n$ cycles, we
get \be |\psi_{\vec k} (nT)\rangle = U^n({\vec k},g_i,g_f)
|\psi_{\vec k}(0) \rangle. \ee For the square pulse case, it is easy
to see that \be  U({\vec k},g_i,g_f) = \exp
\left(-iH_{\vec{k}}(g_f)\frac{T}{2} \right)\exp
\left(-iH_{\vec{k}}(g_i)\frac{T}{2} \right) \label{sqp1}\ee where
$H_{\vec{k}}(g) = (g-b_{\vec{k}})\tau_3+\Delta_{\vec{k}}\tau_1$ as
given in Eq.\ \ref{hamdef1s} in the main text.

\section{Protocol independence}
\label{proin}

We have used the square pulse protocol and the periodic kick 
protocol for the numerical
calculations shown in the main text, mainly due to their simple
analytic forms. However, most of our results are protocol independent
and we illustrate that here with some results using both the linear
ramp and sinusoidal protocols, which we will define below. For
notational simplicity, we restrict ourselves to the 1D Ising model.
\begin{figure}
{\includegraphics[width=\hsize]{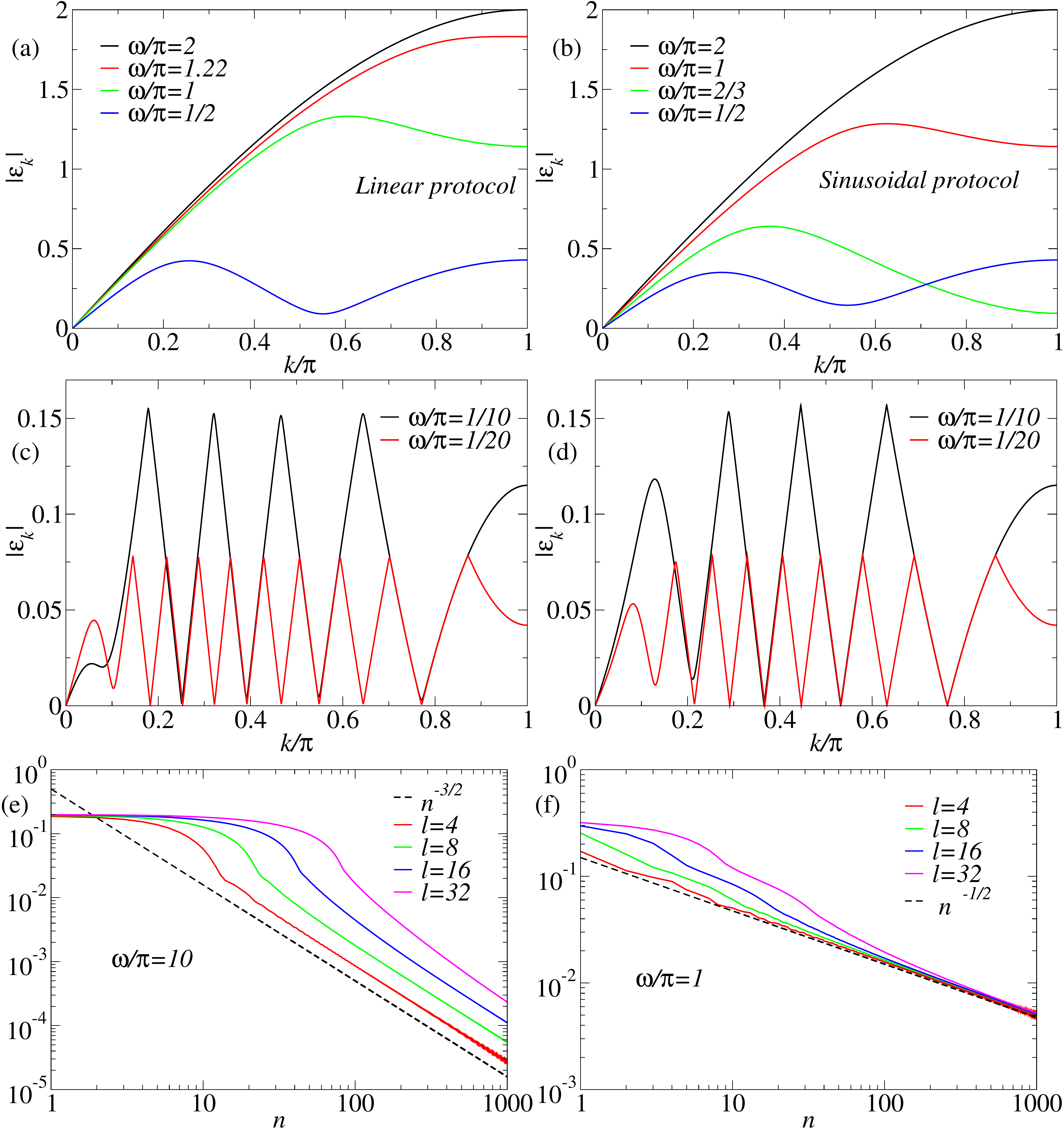}} \caption{Panels (a),(b)
show the behavior of $|\epsilon(k)|$ as a function of $k$ for a few
representative values of $\omega$ both for the linear ramp and the
sinusoidal drive protocols. Panels (c),(d) show $|\epsilon(k)|$ as a
function of $k$ for small values of $\omega$ for the linear ramp and
the sinusoidal drive protocols, respectively. Panels (e),(f) show
the power law decay of $\mathcal{D}$ as a function of $n$ for
several $l$ for the linear ramp protocol. \label{supp2}}
\end{figure}

In the linear ramp protocol with a time period $T=2T_0$, the precise
variation of $g(t)$ between $n$ and $n-1$ cycles is given by
\begin{eqnarray}
g(t) &=& g_i +(g_f-g_i) (t-2(n-1)T_0)/T_0, \nonumber\\
&& {\rm for} \quad 2(n-1)T_0 \le t\le (2n-1)T_0 \nonumber\\
&=& g_f -(g_f-g_i)(t-(2n-1)T_0)/T_0 \nonumber\\
&& {\rm for} \quad (2n-1)T_0 \le t\le 2n T_0 \label{protocol}
\end{eqnarray}
The advantage of this protocol is that one can again obtain exact
analytical solution for the wavefunction at the end of a drive cycle
like the square pulse case, though the solution is more complicated.
The unitary matrix for the evolution of the wavefunction at the end
of one drive can be written as \be
 U_{k}(g_i,g_f) = U_b(k,g_f \rightarrow g_i, T_0)U_f(k,g_i \rightarrow g_f, T_0)
\ee where $U_f$ ($U_b$) refer to the corresponding unitary matrix
for the ``forward'' (``backward'') ramp from $g_i$ to $g_f$ ($g_f$
to $g_i$). Below, we give the explicit expressions for the two
unitary matrices assuming that $g_i > g_f$ without loss of
generality (for details, we refer the reader to Ref.~\onlinecite{vitanov}). 
The matrices can be more easily expressed
through the redefined variables $v = (g_i-g_f)/T_0, T_i
=(b_k-g_i)/\sqrt{v}, T_f = \sqrt{v}T_0-T_i$ and $\omega =
\Delta_k/\sqrt{v}$.

Then, for the forward ramp, we have \be
(U_f)_{11} &=& (U_f)_{22}^* \nonumber \\
&=& \frac{\Gamma(1-i\omega^2/2)}{\sqrt{2 \pi}}[D_{i\omega^2/2}(T_f\sqrt{2}e^{-i\pi/4}) \nonumber \\
&\times& D_{-1+i\omega^2/2}(T_i\sqrt{2}e^{i3\pi/4}) \nonumber \\
&+& D_{i\omega^2/2}(T_f\sqrt{2}e^{i3\pi/4}) \nonumber \\
&\times& D_{-1+i\omega^2/2}(T_i\sqrt{2}e^{-i\pi/4})], \nonumber \\
(U_f)_{12} &=& -(U_f)_{21}^* \nonumber \\
&=& \frac{\Gamma(1-i\omega^2/2)}{\omega \sqrt{\pi}} e^{i\pi/4}[-D_{i\omega^2/2}(T_f\sqrt{2}e^{-i\pi/4}) \nonumber \\
&\times& D_{i\omega^2/2}(T_i\sqrt{2} e^{i3\pi/4}) \nonumber \\
&+&  D_{i\omega^2/2}(T_f\sqrt{2}e^{i3\pi/4})  D_{i\omega^2/2}(T_i\sqrt{2}e^{-i\pi/4})]. \nonumber \\
\ee where $D$ denotes the parabolic cylinder function and $\Gamma$
denotes the gamma function. The unitary matrix for the backward ramp
is then obtained from the above matrix using $(U_b)_{ij} =
(-1)^{i+j}(U_f^*)_{ij}$.

For the sinusoidal protocol with a time period of $T$, the variation
of $g(t)$ is chosen as follows \be g(t) = g_{\rm av}+A \cos
\left(\frac{2\pi t}{T} \right) \ee Here, the unitary matrix for one
drive cycle cannot be expressed analytically and one has to resort
to a numerical solution of the time-dependent Schr\"odinger
equation.

In Fig.~\ref{supp2}, we show the results for the linear ramp
protocol with $g_i=2$, $g_f=0$ and varying $\omega=2\pi/T$ and for
the sinusoidal protocol with $g_{\rm av}=1$, $A=1$ and varying
$\omega=2\pi/T$. From Fig.~\ref{supp2}(a),(b), it is clear that $m$,
the number of zeroes in $d|\epsilon_k|/dk$ for $0<k<\pi$ or
equivalently the number of local extrema in $|\epsilon_k|$ for
$0<k<\pi$, is zero when $\omega \gg 1$ and attains a non-zero value
only below a critical (protocol dependent) $\omega_c$ exactly like
in the square pulse case. From Fig.~\ref{supp2}(c),(d), it is also
clear that $m \sim 1/\omega$ for small $\omega$ for both the linear
ramp and sinusoidal protocols, again like in the case of the square
pulse protocol. Lastly, in Fig.~\ref{supp2}(e),(f), we show the
calculation for the power law decay of $\mathcal{D}$ for two
different $\omega$ on either side of $\omega_c$ for the linear ramp
protocol.

These calculations substantiate the protocol independence of our
results asserted in the main text.

\section{Computation of $S_n(l)$}
\label{pesca}

As shown in the main text, the state generated after $n$ drive
cycles, $|\psi(nT) \rangle$ is the ground state of $\mathcal{H}_t$
(Eq.\ \ref{ht1} of the main text) which is quadratic in the fermionic
operators $c$ and $c^{\dagger}$. For such free fermion Hamiltonians,
the reduced density matrix for the ``ground state'' $|\psi(nT)
\rangle$ can be written as \be \rho_\alpha &=&
\exp(-\mathcal{H}_\alpha)/Z \quad , \mathcal{H}_\alpha =
\sum_{i=1}^l \epsilon_i \eta_i^\dagger \eta_i \label{reducedr} \ee
where $l$ is the number of sites in the subsystem denoted by
$\alpha$ and the operators $\eta_i,\eta_i^\dagger$ are fermionic
operators for single particle states with energies $\epsilon_i$. The
constant $Z$ ensures the correct normalization $tr(\rho_\alpha)=1$.
Since {\it all} correlation functions of the subsystem can be
expressed in terms of the quadratic correlations by using Wick's
theorem, the entanglement Hamiltonian $\mathcal{H}_\alpha$ (and
hence $S(l)$) is determined by the condition that it gives the {\it
right} quadratic correlation functions $C_{ij}$ and $F_{ij}$ for the
sites that belong to the subsystem.

Let us denote the Bogoluibov transformation that gives the diagonal
representation ${\eta,\eta^\dagger}$ from ${c,c^\dagger}$ for the
subsystem as \be
\eta_k &=& \sum_i (g_{ki}c_i + h_{ki}c^\dagger_i) \nonumber \\
\eta_k^\dagger &=& \sum_i (g^*_{ki}c_i^\dagger + h^*_{ki}c_i)
\label{ft} \ee where $i$ belongs to the sites in the subsystem being
considered. Since $\eta,\eta^\dagger$ satisfy anti commutation
algebra, we can easily verify that the matrix $\mathbf{T}$ defined
as

\be \quad
\begin{pmatrix}
\mathbf{g} & \mathbf{h} \\
\mathbf{h}^* & \mathbf{g}^*
\end{pmatrix}
\ee is a $2l \times 2l$ unitary matrix.  Introducing the bra--ket
notation, \be |\phi \rangle =
\begin{pmatrix}
c \\
c^\dagger
\end{pmatrix}
\ee and \be |\psi \rangle =
\begin{pmatrix}
\eta  \\
\eta^\dagger
\end{pmatrix}
\ee The transformation in Eq.~\ref{ft} is then simply expressed as
\be |\psi \rangle  = \mathbf{T} |\phi \rangle \ee Expressing the
entanglement Hamiltonian $\mathcal{H}_\alpha$ as $\phi^\dagger
\mathbf{M} \phi$, we see that \be \mathbf{M} = \mathbf{T}^\dagger
\begin{pmatrix}
\epsilon & 0  \\
0 & -\epsilon
\end{pmatrix}
 \mathbf{T}
\label{Hd} \ee where the middle matrix is diagonal. Now, calculating
the outer product of $|\phi \rangle$ with itself, which we denote by
the $2l \times 2l$ matrix $\mathcal{C}$ and which requires the
knowledge of two $l \times l$ matrices $\mathbf{C}$ and
$\mathbf{F}$, we get \be
\mathcal{C}=\langle |\phi \rangle \langle \phi| \rangle= \mathbf{T}^\dagger \langle |\psi \rangle \langle \psi| \rangle \mathbf{T} \nonumber \\
\mathbf{T}^\dagger
\begin{pmatrix}
\frac{1}{\exp(-\epsilon)+1} & 0  \\
0 &  \frac{1}{\exp(\epsilon)+1}
\end{pmatrix}
 \mathbf{T}
\ee where the middle matrix is again diagonal and we have used
$\langle \eta^\dagger_k \eta_{k'} \rangle = \frac{1}
{\exp(\epsilon_k)+1} \delta_{k,k'}$. The eigenvalues of
$\mathcal{C}$ always come in pairs $p_k,1-p_k$ where $p_k$ is the
probability of occupation of the $k^{th}$ fermionic modes. Then the
entanglement entropy $S(l)$ is simply $-\sum_{i=1}^{2l} p_i
\log(p_i)$. This result has been used in numerical calculations of
the main text.

\section{Details of calculation of ${\mathcal D}$}
\label{secmeasure}

In this section, we study the relaxation of the system for finite
but large $n$ to its diagonal ensemble (equivalently GGE) value
discussed in the past section. To this end, we define a distance
measure which provides us information regarding this relaxation as a
function of $n$. For the class of integrable models we consider
here, the entanglement properties are solely determined by the
two-point correlations of the subsystem as shown in the past
section. Thus we define the distance measure using the correlation
matrices $\mathcal{C}_\infty(l)$ and $\mathcal{C}_n(l)$. This can be
done by using the standard trace distance between these two matrices
\cite{essler1}:
\begin{equation}
D(\mathcal{C}_n(l),\mathcal{C}_\infty(l)) = \frac{1}{2l}
Tr\sqrt{(\mathcal{C}_\infty(l)-\mathcal{C}_n(l))^\dagger
(\mathcal{C}_\infty(l)-\mathcal{C}_n(l))}
\end{equation}
We note that $D$ is positive, lies in $\in [0,1]$, and equals zero
only when the two matrices are identical. The matrix elements of
$\mathcal{C}_n(l)$ are determined by $\langle c^{\dagger}_{\vec{i}}
c_{\vec{j}} \rangle$ and $\langle c^{\dagger}_{\vec{i}}
c^{\dagger}_{\vec{j}} \rangle$; so we first calculate how these
quantities behave as a function of $n$ when $n \gg 1$. Using Eq.\
\ref{ueq3}, we find, after a few lines of algebra, that in the
thermodynamic limit ($L \rightarrow \infty$),
\begin{widetext}
\begin{eqnarray}
\langle c_{\vec i}^{\dagger} c_{\vec j} \rangle &=& \frac{2}{(2\pi)^d} \int_{\vec{k} \in BZ/2} d^dk \ \cos(\vec{k}\cdot(\vec{i}-\vec{j}))\left (\frac{u_{0\vec{k}}^2}{2}(1+\hat{n}_{\vec{k}3}^2) +\frac{v_{0\vec{k}}^2}{2}(1-\hat{n}_{\vec{k}3}^2) + u_{0\vec{k}}v_{0\vec{k}}\hat{n}_{\vec{k}1}\hat{n}_{\vec{k}3}\right) \rightarrow \mathrm{GGE} \nonumber \\
&+& \frac{2}{(2\pi)^d} \int_{\vec{k} \in BZ/2} d^dk \cos(\vec{k}\cdot(\vec{i}-\vec{j}))\left ( \left(\frac{u_{0\vec{k}}^2}{2}-\frac{v_{0\vec{k}}^2}{2}\right)(1-\hat{n}_{\vec{k}3}^2)- u_{0\vec{k}}v_{0\vec{k}}\hat{n}_{\vec{k}1}\hat{n}_{\vec{k}3}\right)\cos(2n\phi_{\vec k}) \nonumber \\
&-&\frac{2}{(2\pi)^d} \int_{\vec{k} \in BZ/2} d^dk \cos(\vec{k}\cdot(\vec{i}-\vec{j})) u_{0\vec{k}}v_{0\vec{k}}\hat{n}_{\vec{k}2}\sin(2n\phi_{\vec k}) \nonumber \\
\langle c_{\vec{i}}^{\dagger} c^{\dagger}_{\vec{j}} \rangle &=& \frac{2}{(2\pi)^d} \int_{\vec{k} \in BZ/2} d^dk \sin(\vec{k}\cdot(\vec{i}-\vec{j})) \left(u_{0\vec{k}}v_{0\vec{k}}\hat{n}_{\vec{k}1}(\hat{n}_{\vec{k}1}+i\hat{n}_{\vec{k}2}) + \frac{(u_{0\vec{k}}^2-v_{0\vec{k}}^2)\hat{n}_{\vec{k}3}(\hat{n}_{\vec{k}1}+i\hat{n}_{\vec{k}2})}{2}\right)\rightarrow \mathrm{GGE} \nonumber \\
&+&\frac{2}{(2\pi)^d} \int_{\vec{k} \in BZ/2} d^dk  \sin(\vec{k}\cdot(\vec{i}-\vec{j})) \left(u_{0\vec{k}}v_{0\vec{k}}(1-\hat{n}_{\vec{k}1}^2-i\hat{n}_{\vec{k}1}\hat{n}_{\vec{k}2}) -\frac{(u_{0\vec{k}}^2-v_{0\vec{k}}^2)\hat{n}_{\vec{k}3}(\hat{n}_{\vec{k}1}+i\hat{n}_{\vec{k}2})}{2}\right) \cos(2n \phi_{\vec k}) \nonumber \\
&+&  \frac{2}{(2\pi)^d} \int_{\vec{k} \in BZ/2} d^dk
\sin(\vec{k}\cdot(\vec{i}-\vec{j}))
\left(iu_{0\vec{k}}v_{0\vec{k}}\hat{n}_{\vec{k}3}-i\frac{(u_{0\vec{k}}^2-v_{0\vec{k}}^2)(\hat{n}_{\vec{k}1}+i\hat{n}_{\vec{k}2})}{2}\right)
\sin(2n \phi_{\vec k}) \label{generalresult}
\end{eqnarray}
\end{widetext}
where the integral is taken over half the Brillouin zone (BZ) since
the $(\vec{k},-\vec{k})$ fermions are always excited in pairs. It is
clear from Eq.\ \ref{generalresult} that only the terms indicated by
GGE survive in the $n \rightarrow \infty$ limit. These have been
represented as $\langle c_i^{\dagger} c_j \rangle_{\infty}$ and $
\langle c_i c_j \rangle_{\infty}$ in the main text. The other terms
lead to Eq.\ \ref{ffns} of the main text for $u_{\vec k}^i \equiv u_{\vec k
0} =0$ and $v_{\vec k i}\equiv v_{\vec k 0}=1$.

Before ending this section, we note that for large $\omega$ in
general the condition $\sin(\theta_{\vec k})=0 = d\alpha_{\vec
k}/dk_i$ is expected to be satisfied at the minima of $H_{\vec k}$.
If these minima happen to be at the zone boundary, where $\sin( \vec
k \cdot (\vec i -\vec j))$ vanishes, then $f(\vec k)=0$ (where
$f(\vec k)$ can be read off from Eq.\ \ref{generalresult}), and
hence the relaxation of $S$ to GGE will scale as
$(\omega/n)^{(d+2)/2}$. However, if the minima of $H_{\vec k}$
occurs at $\vec k= \vec k_1$ so that $\sin( \vec k_1 \cdot (\vec i
-\vec j)) \ne 0$, then $f(\vec{k}_1) \ne 0$ for $\langle
c_{\vec{i}}^{\dagger} c^{\dagger}_{\vec{j}}\rangle$ and hence $S$
does not scale as $(\omega/n)^{(d+2)/2}$. We note however, that even
in this case, all local quantities (such as fermion density which
corresponds to magnetization in the spin language) and those which
depend only on diagonal correlation functions $\langle c_i^{\dagger}
c_j\rangle$ still exhibits $(\omega/n)^{(d+2)/2}$ scaling and will
show the two dynamical regimes discussed in the main text.

\section{Construction of $H_F$}
\label{flcons}

Let us consider an arbitrary periodic protocol characterized by the
number of cycles $n$ and the drive frequency $\omega= 2\pi/T$ which
takes the system from an initial state $\psi^i =\prod_{\vec k}
\psi_{\vec k}^i = \prod_{\vec k} (u_{\vec k}^i,v_{\vec k}^i)^T$ to a
final state $\psi^f =\prod_{\vec k} \psi_{\vec k}^f = \prod_{\vec k}
(u_{\vec k}^{nf},v_{\vec k}^{nf})^T$. In what follows we shall
define the state reached after one drive cycle to be $\psi'
=\prod_{\vec k} \psi'_{\vec k} = \prod_{\vec k} (u_{\vec
k}^{f},v_{\vec k}^{f})^T$. One can relate the wavefunctions
$\psi'_{\vec k}$ and $\psi^i_{\vec k}$ through a evolution operator
$U_{\vec k}$ given by
\begin{eqnarray}
\label{ueq1} \psi_{\vec k}^f &=& U_{\vec k}^n \psi_{\vec k}^i, \quad
\psi'_{\vec k} = U_{\vec k} \psi_{\vec k}^i,
\\
U_{\vec k} &=& \left(\begin{array}{cc}
\cos(\theta_{\vec k}) e^{i \alpha_{\vec k}} & \sin(\theta_{\vec k})e^{i \gamma_{\vec k}} \\
-\sin(\theta_{\vec k})e^{-i \gamma_{\vec k}}& \cos(\theta_{\vec k}) e^{-i \alpha_{\vec k}} \\
\end{array} \right)
= e^{-i H_{\vec{k}F} T}  \nonumber
\end{eqnarray}
The parametrization of $U_{\vec k}$ follows from unitary nature of
the evolution and $\theta_{\vec k}$, $\alpha_{\vec k}$ and
$\gamma_{\vec k}$ are real-valued functions of $\vec k$. Here $H_F$
is the Floquet Hamiltonian which can be used to describe the state
of the system after a drive period; note that the final state after
$n$ periods is simply described by $ \exp[-i n H_{\vec{k}F} T]$.

To make further progress, we express $U$ in terms of $\psi^i_{\vec
k}$ and $\psi'_{\vec k}$. A few lines of straightforward algebra
yields
\begin{eqnarray}
\label{ueq2} \sin^2(\theta_{\vec k}) &=& \left[|u_{\vec k}^f|^2
v_{\vec k}^{i2} + |v_{\vec k}^f|^2 u_{\vec k}^{i2}
\right. \nonumber\\
&& \left. - 2 |u_{\vec k}^f| |v_{\vec k}^f| u_{\vec k}^{i} v_{\vec
k}^{i} \cos(\mu_{\vec k} - \mu'_{\vec k})
\right] \\
\gamma_{\vec k} &=& \arctan \left( \frac{ |u_{\vec k}^f| v_{\vec
k}^i \sin(\mu_{\vec k}) + u_{\vec k}^{i} |v_{\vec k}^f|
\sin(\mu'_{\vec k})}{|u_{\vec k}^f| v_{\vec k}^i \cos(\mu_{\vec k})
- u_{\vec k}^{i}
|v_{\vec k}^f| \cos(\mu'_{\vec k})} \right) \nonumber\\
\alpha_{\vec k} &=&  \arctan \left( \frac{ |u_{\vec k}^f| u_{\vec
k}^i \sin(\mu_{\vec k}) - v_{\vec k}^{i} |v_{\vec k}^f|
\sin(\mu'_{\vec k})}{|u_{\vec k}^f| u_{\vec k}^i \cos(\mu_{\vec k})
+ |v_{\vec k}^f| v_{\vec k}^i \cos(\mu'_{\vec k})} \right) \nonumber
\end{eqnarray}
where we have taken $u_{\vec k}^i$ and $v_{\vec k}^i$ to be real and
have parameterized $u_{\vec k}^f = |u_{\vec k}^f| \exp[i \mu_{\vec
k}]$ and $ v_{\vec k}^f = |v_{\vec k}^f| \exp[i \mu'_{\vec k}]$. We
note that $U_{\vec k}$ reduces to the identity matrix for $u_{\vec
k}^f=u_{\vec k}^i$ and $v_{\vec k}^f= v_{\vec k}^i$. Note that for
$u_{\vec k}^i=0$ and $v_{\vec k}^i=1$, Eq.\ \ref{ueq2} reduces to
the expressions used in the main text.

Next, we obtain an expression for $H_{\vec{k}F}$. To do this, we
note that in the present case $H_{\vec{k}F}$ can be written in the
form of a $2 \times 2$ matrix which can be expressed in terms of
Pauli matrices: $H_{\vec k F}= \vec \sigma \cdot \vec \epsilon_{\vec
k}$. where $\vec \epsilon_{\vec k} = (\epsilon_{1k}, \epsilon_{2k},
\epsilon_{3k})$. This allows us to write
\begin{eqnarray}
\label{ueq3} U_{\vec k} &=& e^{-i (\vec \sigma \cdot \vec n_{\vec
k}) \phi_{\vec k}}, \quad n_{\vec k}= \frac{\vec \epsilon_{\vec
k}}{|\vec \epsilon_{\vec k}|} \quad, \phi_{\vec k} = T|\vec
\epsilon_{\vec k}|
\end{eqnarray}
Using Eqs.\ \ref{ueq1}, \ref{ueq2}, and \ref{ueq3}, one obtains,
after some straightforward algebra,
\begin{eqnarray}
n_{\vec{k}1} &=& -\sin(\theta_{\vec k}) \sin(\gamma_{\vec k})\sin(\phi_{\vec k})/D_{\vec k}
\nonumber\\
n_{\vec{k}2} &=& -\sin(\theta_{\vec k}) \cos(\gamma_{\vec k})\sin(\phi_{\vec k})/D_{\vec k}
\nonumber\\
n_{\vec{k}3} &=& -\cos(\theta_{\vec k}) \sin(\alpha_{\vec k})\sin(\phi_{\vec k})/D_{\vec k}
\nonumber\\
D_{\vec k}&=& \sqrt{1-\cos^2(\theta_{\vec k}) \cos^2(\alpha_{\vec
k})} \nonumber\\
|\vec \epsilon_{\vec k}| &=&   \arccos[\cos(\theta_{\vec k})
\cos(\alpha_{\vec k})]/T.  \label{ueq3a}
\end{eqnarray}
The last of these
expressions is Eq.\ \ref{ueq3as} of the main text. At the edge of the Brillouin
zone, where the off-diagonal component of $H_k $ disappears, $U_k$
becomes a diagonal matrix, which in turns makes $\sin(\theta_{\vec k
})=0$. This leads us to the result $n_{\vec{k}1}= n_{\vec{k}2}=0$
and $n_{\vec{k}3}=\pm 1$ for these momenta values which is used in
the main text.

Next, we  provide an explicit expression for $|\epsilon_{\vec k}|$
for the square pulse protocol defined in the main text. This can be
done by combining Eq.\ \ref{sqp1} with Eqs.\ \ref{ueq2}, \ref{ueq3}
and \ref{ueq3a}. A somewhat lengthy calculation yields \be
|\epsilon_{\vec k}| = \arccos(M_{\vec k})/T \ee with \be
M_{\vec k} =  \cos(\Phi_{\vec k i} )\cos(\Phi_{\vec k f}) - \hat{N}_{\vec k i}
\cdot \hat{N}_{\vec k f} \sin(\Phi_{\vec k i})\sin(\Phi_{\vec k f}) \nonumber \\
\ee where $\Phi_{\vec k i(f)}=E_{\vec k i(f)} T/2$ with $E_{\vec k
i(f)} = \sqrt{(g_{i(f)}-b_{\vec k})^2+\Delta_{\vec k}^2}$ and the
components of $\hat N$ are given by
\begin{eqnarray}
\hat{N}_{\vec k i(f)} = \left(\frac{\Delta_{\vec k}}{E_{\vec k
i(f)}},0,\frac{g_{i(f)}-b_{\vec k}}{E_{\vec k i(f)}} \right).
\end{eqnarray}

We now show how to compute $\omega_c$, i.e. the first critical drive frequency 
for the dynamical phase transition as the drive frequency $\omega$ is 
reduced from $\omega \gg 1$ for this protocol. For
simplicity, we restrict to the one dimensional case where $k \in
[0,\pi]$. First, we note that a new zero in
$d|\epsilon_k|/dk$ can only appear from the boundaries $k=0,\pi$. We
have numerically checked that the appearance of the first extra zero
in $d|\epsilon_k|/dk$ is from $k=\pi$ for the square pulse protocol. Then,
$\omega_c$ can be simply calculated by expanding $d|\epsilon_k|/dk$
for $k=\pi-\epsilon$ and finding the value of $\omega$ where the
$\mathcal{O}(\epsilon)$ term first changes its sign. This leads to
Eq.\ \ref{condtrans} of the main text.


\begin{thebibliography}{99}
\bibitem{comment2d} For 2D Kitaev model on a torus, $S^{\rm 2D}_n(l) = \sum_{k_y} S_n(l;k_y)$
where we have chosen the subsystem as in Ref.\ \cite{yao1}. Each of
these $S_n(l;k_y)$ is expected to behave like $S_n(l)$ for 1D system
[Fig.\ \ref{fig1}(c)]. Thus the non-area, non-volume scaling of
$S^{\rm 2D}_n(l)$ is expected to hold for 2D systems as well.
\bibitem{comment4} We note that similar features occur for sinusodal and
linear ramp protocols as shown in App.\ \ref{proin}.
\end{thebibliography}
\end{document}